\documentclass[twocolumn,twocolappendix]{aastex631}    

\usepackage{booktabs}
\usepackage{longtable}
\usepackage{appendix}
\usepackage{amsmath}
\usepackage{amssymb}
\usepackage{graphicx}
\usepackage{mfirstuc}
\usepackage{rotating}
\usepackage{soul}
\usepackage[encapsulated]{CJK}
\usepackage[utf8]{inputenc}

\newcommand{\db}{$\delta_\textrm{burst}$}
\newcommand{\sigsf}{$A_\textrm{SFMS}$}
\newcommand{\ha}{H$\alpha$}
\newcommand{\haew}{H$\alpha$~EW}

\graphicspath{{figures/, tables/}}
\shorttitle{Long Burst Cycles in Low-mass Galaxies at Cosmic Noon}
\shortauthors{Mintz et al.}

\begin{document}

\title{Taking a Break at Cosmic Noon: Continuum-selected Low-mass Galaxies Require Long Burst Cycles}

\author[0000-0002-9816-9300]{Abby Mintz}
\affil{Department of Astrophysical Sciences, Princeton University, 4 Ivy Lane, Princeton, NJ 08544, USA}

\author[0000-0003-4075-7393]{David J. Setton}\thanks{Brinson Prize Fellow}
\affiliation{Department of Astrophysical Sciences, Princeton University, 4 Ivy Lane, Princeton, NJ 08544, USA}

\author[0000-0002-5612-3427]{Jenny E. Greene}
\affiliation{Department of Astrophysical Sciences, Princeton University, 4 Ivy Lane, Princeton, NJ 08544, USA}

\author[0000-0001-6755-1315]{Joel Leja}
\affiliation{Department of Astronomy \& Astrophysics, The Pennsylvania State University, University Park, PA 16802, USA}
\affiliation{Institute for Computational \& Data Sciences, The Pennsylvania State University, University Park, PA 16802, USA}
\affiliation{Institute for Gravitation and the Cosmos, The Pennsylvania State University, University Park, PA 16802, USA}

\author[0000-0001-9269-5046]{Bingjie Wang (\begin{CJK*}{UTF8}{gbsn}王冰洁\ignorespacesafterend\end{CJK*})}
\affiliation{Department of Astronomy \& Astrophysics, The Pennsylvania State University, University Park, PA 16802, USA}
\affiliation{Institute for Computational \& Data Sciences, The Pennsylvania State University, University Park, PA 16802, USA}
\affiliation{Institute for Gravitation and the Cosmos, The Pennsylvania State University, University Park, PA 16802, USA}

\author[0000-0001-8174-317X]{Emilie Burnham}
\affiliation{Department of Astronomy \& Astrophysics, The Pennsylvania State University, University Park, PA 16802, USA}
\affiliation{Institute for Computational \& Data Sciences, The Pennsylvania State University, University Park, PA 16802, USA}

\author[0000-0002-1714-1905]{Katherine A. Suess}
\affiliation{Department for Astrophysical and Planetary Science, University of Colorado, Boulder, CO 80309, USA}

\author[0000-0002-7570-0824]{Hakim Atek}
\affiliation{Institut d’Astrophysique de Paris, CNRS, Sorbonne Universit\'e, 98bis Boulevard Arago, 75014, Paris, France}

\author[0000-0001-5063-8254]{Rachel Bezanson}
\affiliation{Department of Physics and Astronomy and PITT PACC, University of Pittsburgh, Pittsburgh, PA 15260, USA}

\author[0000-0001-6755-1315]{Gabriel Brammer}
\affiliation{Cosmic Dawn Center (DAWN), Copenhagen, Denmark}
\affiliation{Niels Bohr Institute, University of Copenhagen, Jagtvej 128, Copenhagen, Denmark}

\author[0000-0002-7031-2865]{Sam E. Cutler}
\affiliation{Department of Astronomy, University of Massachusetts, Amherst, MA 01003, USA}

\author[0000-0001-8460-1564]{Pratika Dayal}
\affiliation{Kapteyn Astronomical Institute, University of Groningen, P.O. Box 800, 9700 AV Groningen, The Netherlands}

\author[0000-0002-1109-1919]{Robert Feldmann}
\affiliation{Department of Astrophysics, Universität Zürich, Winterthurerstrasse 190, Zürich, CH-8057, Switzerland}

\author[0000-0001-6278-032X]{Lukas J. Furtak}
\affiliation{Physics Department, Ben-Gurion University of the Negev, P.O. Box 653, Be’er-Sheva 84105, Israel}

\author[0000-0002-3254-9044]{Karl Glazebrook}
\affiliation{Centre for Astrophysics and Supercomputing, Swinburne University of Technology, PO Box 218, Hawthorn, VIC 3122, Australia}

\author[0000-0002-3475-7648]{Gourav Khullar}
\affiliation{Department of Astronomy, University of Washington, Physics-Astronomy Building, Box 351580, Seattle, WA 98195-1700, USA}
\affiliation{eScience Institute, University of Washington, Physics-Astronomy Building, Box 351580, Seattle, WA 98195-1700, USA}

\author[0000-0002-5588-9156]{Vasily Kokorev}
\affiliation{Department of Astronomy, The University of Texas at Austin, Austin, TX 78712, USA}

\author[0000-0002-2057-5376]{Ivo Labb\'e}
\affiliation{Centre for Astrophysics and Supercomputing, Swinburne University of Technology, Melbourne, VIC 3122, Australia}

\author[0000-0003-2871-127X]{Jorryt Matthee}
\affiliation{Institute of Science and Technology Austria (ISTA), Am Campus 1, 3400 Klosterneuburg, Austria}

\author[0000-0003-0695-4414]{Michael V. Maseda}
\affiliation{Department of Astronomy, University of Wisconsin-Madison, Madison, WI 53706, USA}

\author[0000-0001-8367-6265]{Tim B. Miller}
\affiliation{Center for Interdisciplinary Exploration and Research in Astrophysics (CIERA), Northwestern University,1800 Sherman Ave, Evanston, IL 60201, USA}

\author[0000-0001-7300-9450]{Ikki Mitsuhashi}
\affiliation{Department for Astrophysical and Planetary Science, University of Colorado, Boulder, CO 80309, USA}

\author[0000-0003-2804-0648 ]{Themiya Nanayakkara}
\affiliation{Centre for Astrophysics and Supercomputing, Swinburne University of Technology, PO Box 218, Hawthorn, VIC 3122, Australia}

\author[0000-0002-9651-5716]{Richard Pan}
\affiliation{Department of Physics \& Astronomy, Tufts University, MA 02155, USA}

\author[0000-0002-0108-4176]{Sedona H. Price}
\affiliation{Space Telescope Science Institute (STScI), 3700 San Martin Drive, Baltimore, MD 21218, USA}

\author[0000-0003-1614-196X]{John R. Weaver}
\affiliation{Department of Astronomy, University of Massachusetts, Amherst, MA 01003, USA}

\author[0000-0001-7160-3632]{Katherine E. Whitaker}
\affiliation{Department of Astronomy, University of Massachusetts, Amherst, MA 01003, USA}
\affiliation{Cosmic Dawn Center (DAWN), Copenhagen, Denmark}

\author[0000-0001-7160-3632]{Belinda Wu}
\affiliation{Department of Astrophysical Sciences, Princeton University, 4 Ivy Lane, Princeton, NJ 08544, USA}

\begin{abstract}
While bursty star formation in low-mass galaxies has been observed in local populations and reproduced in simulations, the dormant phase of the burst cycle has not been well studied beyond the local Universe due to observational limitations. We present a unique sample of 43 JWST PRISM spectra of low-mass galaxies ($M_\star < 10^{9.5}\,M_\odot$) at cosmic noon ($1<z<3$), uniformly selected on F200W magnitude and precise photometric redshifts enabled by 20-band JWST photometry from the UNCOVER and MegaScience surveys. The spectra reveal numerous strong Balmer breaks, which are negatively correlated with the galaxies' \ha\ equivalent width. By comparing these observations to synthetic samples of spectra generated using a simple parametrization of bursty star formation histories, we show that star formation in low-mass galaxies at cosmic noon is likely dominated by burst cycles with long timescales ($\gtrsim 100$ Myr) and large deviations below the star-forming main sequence ($\gtrsim 0.8$ dex). Our results suggest that galaxies in this population--at least those within our detection limits--should not be classified solely by their current star formation rates, but instead viewed as a unified population undergoing dynamic movement above and below the star-forming main sequence. The derived constraints demonstrate that long-timescale fluctuations are important for this class of galaxies, indicating that galaxy-scale gas cycles--rather than molecular-cloud-scale stochasticity--are the primary regulators of star formation variability in low-mass galaxies at cosmic noon. 
\end{abstract}


\section{Introduction} \label{sec:intro}

Recent discoveries of UV-luminous high-redshift galaxies with JWST have led to increased interest in bursty star formation in the early Universe as a means of reconciling unexpected observations with long-held theories of galaxy assembly \citep{Whitler2023, Endsley2024, Looser2025, Cole2025}. But long before the launch of JWST, bursty star formation had been observed, studied, and simulated in low-mass galaxies in the local Universe. Star formation histories (SFHs) derived from color-magnitude diagrams of resolved stellar populations in local dwarfs demonstrated that these low-mass systems do not perfectly follow the typical picture of smooth galaxy evolution and instead experience extended bursts of star formation \citep{Tolstoy2009, McQuinn2010a, McQuinn2010b, Weisz2011}. While most galaxies are observed to fall on a tight relation between stellar mass and star formation rate known as the star-forming main sequence \citep[SFMS,][]{Brinchmann2004, Noeske2007, Whitaker2012, Speagle2014, Leja2022}, lower mass galaxies undergo extreme fluctuations above the SFMS, with some evidence suggesting increased scatter about the relation for $M_\star \lesssim 10^9\, M_\odot$ \citep{Atek2014, Santini2017, Atek2022, Merida2023}.

These early observational hints of burstiness have been further substantiated by simulations, which find increased variability in star formation in low-mass systems when including realistic baryonic feedback \citep{Sparre2017, Orr2017, Feldmann2017b, Cenci2024, Dome2025, Fortune2025}. Feedback, driven by supernovae and stellar winds, is now well-established as a crucial process regulating star formation, morphology, density profiles, and chemical composition of low-mass galaxies \citep{Collins2022}. Given that the production of the massive stars at the center of these feedback processes is directly tied to the specifics of star formation in these systems, understanding bursty star formation – its prevalence in various populations, its timescales and amplitudes – is essential for building a complete picture of baryonic feedback and consequently of galaxy evolution at all masses and redshifts. 

In particular, constraining the timescale of star-formation bursts, sometimes described as the burst duration, provides insight into the underlying physics of the burst-regulating processes \citep{Faucher-Giguere2018, Caplar2019, Tacchella2020, Iyer2020, Iyer2024, Cenci2024}. Short bursts can likely be explained by stochasticity in the formation and destruction of a small number of discrete giant molecular clouds while long-timescale bursts reflect regulation by global gas-cycling, the feedback-driven exchange of material between the galaxy and its circum-galactic medium \citep{Hopkins2023}.

While model uncertainties make it challenging to accurately and precisely recover the SFH of a single galaxy from its observed photometry or spectrum \citep{Wang2025}, population-level constraints on burst parameters have been successfully derived from distributions of SF-tracing observables. The most common approach relies on comparing \ha\ and ultraviolet (UV) emission, which probe star formation on timescales of $\sim$10 and 100 Myr respectively, approximately the main-sequence lifetimes of the luminous O and B stars that dominate the ionizing and UV emission \citep{Kennicutt2012}. The difference in SFRs derived from these two observables for a large number of galaxies should therefore reveal information about the timescale of recent variable star formation \citep{FloresVelazquez2021}. This technique has been employed in the local Universe to estimate typical burst timescales and amplitudes \citep{Glazebrook1999, Weisz2012, Guo2016, Emami2019, Broussard2022} and at high-$z$ to identify the prevalence of burstiness in early galaxy ensembles \citep{Faisst2019, Atek2022}. Though commonly used, the UV-\ha\ comparison method is complicated by uncertainty in attenuation corrections and the degeneracy of signatures caused by IMF variations and burstiness, with some arguing that it is not a reliable tracer \citep{Rezaee2023}.

The Balmer break, a sharp increase in continuum emission at 3645\,\AA\ is more robust to the complications plaguing UV emission, and when measurable, can provide promising constraints on burstiness. Caused by hydrogen absorption of photons with energies of at least 3.4 eV, the Balmer break is most prominent in A-type stars, which have photospheres at the optimal temperature to sustain significant hydrogen in the $n=2$ energy level. In galaxy spectra, the Balmer break is therefore used as an indicator of average stellar age, with a stronger break indicating a larger contribution from A stars over shorter-lived O- and B-type stars. The break strength only develops after star formation has been suppressed for some time and traces star formation on timescales of $\gtrsim100$ Myr. Photometric proxies for the break have been used to study starbursts in more massive galaxies \citep{Nanayakkara2020} and the 4000\,\AA\ break, a nearby dip in emission caused by metal-line blanketing in older stellar populations, has been used to constrain burstiness \citep{Kauffmann2014} at low-$z$. The 4000\,\AA\ break traces star formation on longer timescales ($\gtrsim1$ Gyr) than the Balmer break--as can photometric indices, which cannot distinguish between the two features.

Beyond the Local Volume, observational work on low-mass burstiness has relied on the detection of extreme line-emitters, low-mass systems undergoing episodes of intense star formation \citep{Maseda2013, Maseda2014, Atek2022}. While such samples have revealed extreme variability in low-mass galaxies at cosmic noon, they are limited by their shallow depth and probe only the high-SFR tail of a burst cycle. As such, these studies reveal little about the durations or depth of the burst cycles. The other half of the cycle, the dormant phase when the galaxy drops below the main sequence, has not been comprehensively studied in this mass and redshift regime. However, the recent discovery of galaxies with downturns in star formation at high-$z$ \citep{Strait2023, Alberts2024, Endsley2024, Dome2024, Looser2024, Looser2025, Trussler2025, Takahashi2025} highlights the importance of understanding the dormant phase and of constructing samples that span a range of SFR. Now, with the launch of JWST, we are able to construct magnitude-limited samples beyond the local Universe down to much lower masses than was previously possible, selecting based on rest-optical continuum emission instead of emission lines or bright UV. Low-resolution PRISM spectra can detect the continuum of these galaxies, providing a window into the equally important dormant phase of burst cycles in these populations. 

\begin{figure*}[t]
    \centering
    \includegraphics[width=1\linewidth]{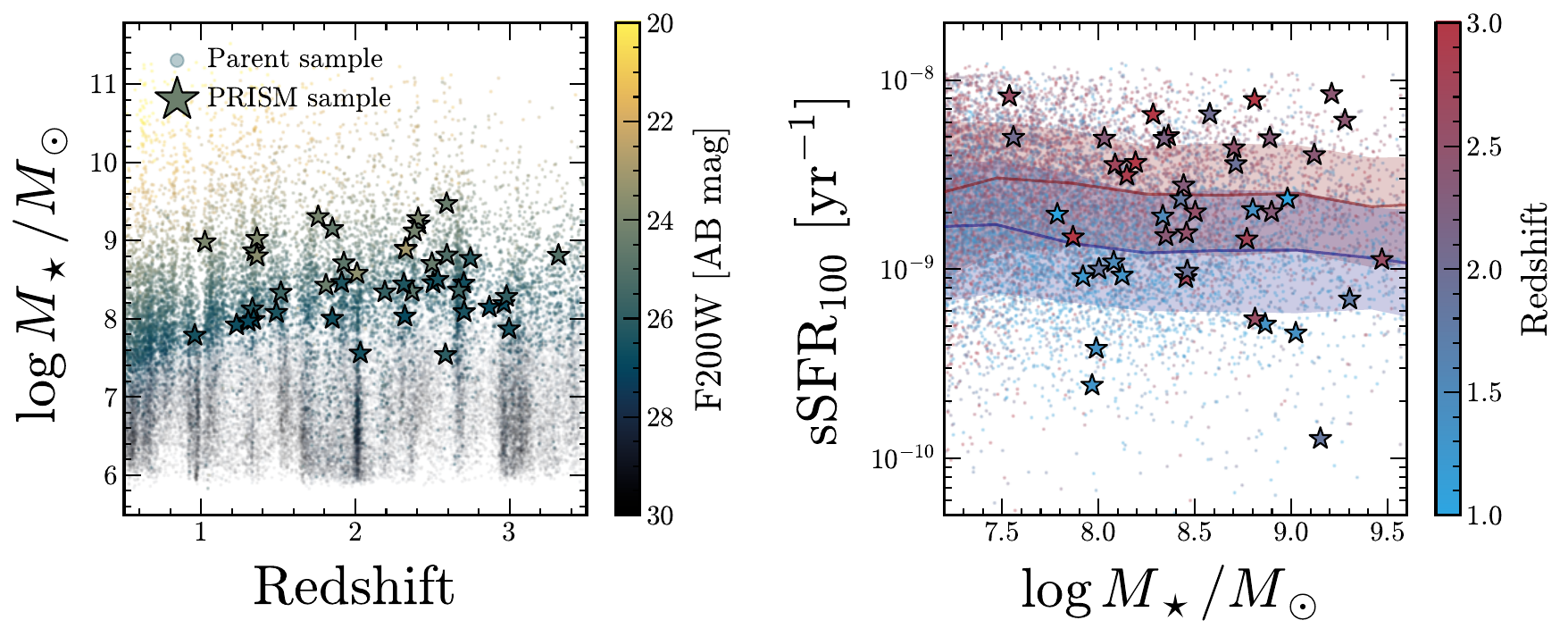}
    \caption{Left: Stellar mass versus redshift for the sources in the PRISM sample and parent sample, colored by their F200W magnitude. The PRISM sources are plotted at the position of their spectroscopic redshift while the parent sample uses photometric redshifts. The PRISM sample was selected on F200W magnitude (F200W $<$ 27) and photometric redshift ($1<z_\text{phot} <3$). Right: Log specific star formation rate versus log stellar mass for the PRISM sample and parent sample, colored by redshift. The solid lines and shaded region show the median sSFR as a function of mass for the parent sample in two redshift bins: $z\in[1,2]$ in blue and $z\in[2,3]$ in red. The PRISM sample spans the sSFR range, with a significant number of sources above and below the inner quartile range.}
    \label{fig:samplesummary}
\end{figure*}

\begin{figure*}[th]
    \centering
    \includegraphics[width=1\linewidth]{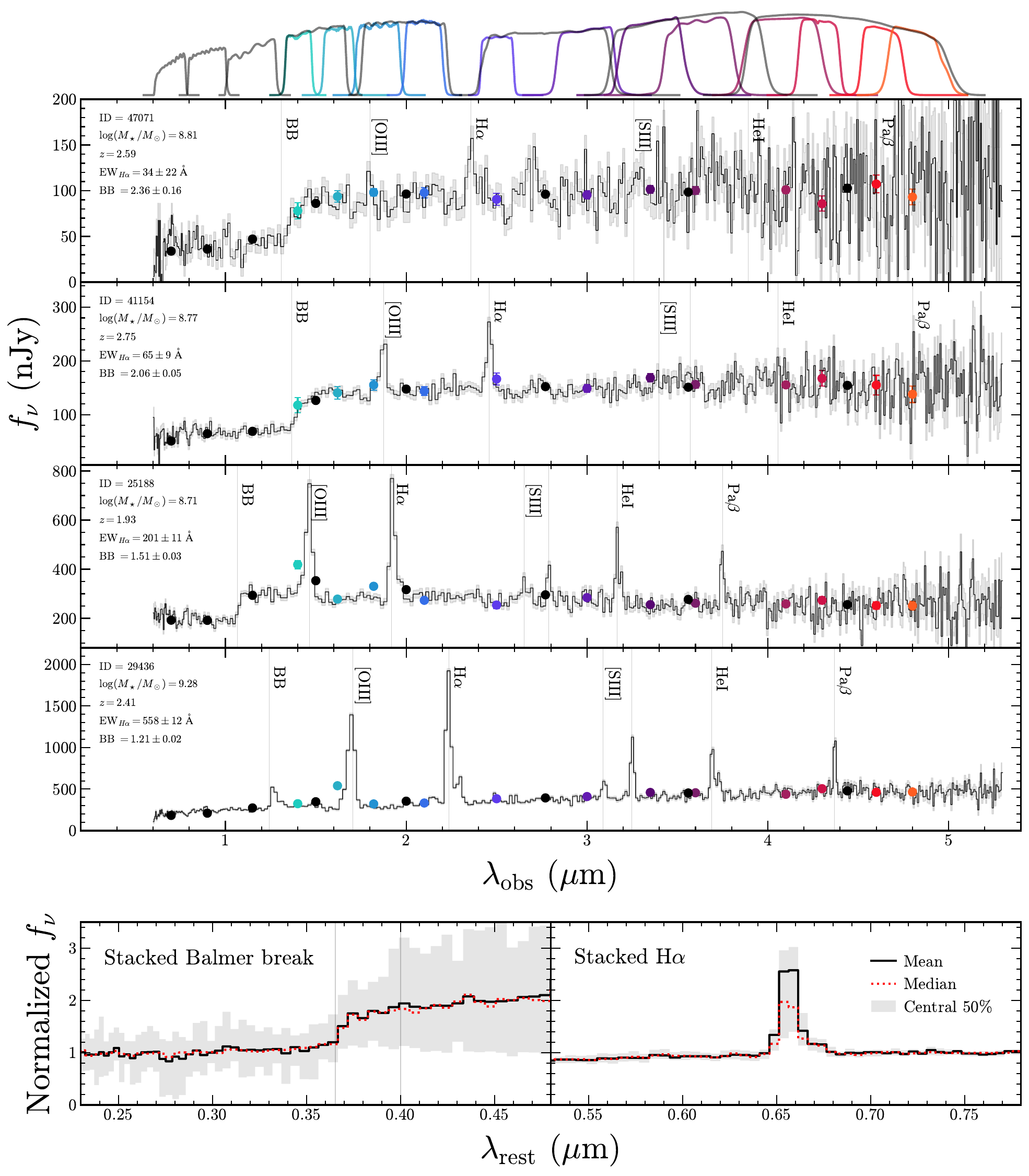}
    \caption{Top: PRISM spectra and MegaScience JWST photometry for four galaxies in our low-mass cosmic noon sample. The galaxies are ordered by decreasing Balmer break strength and increasing \haew. The 20 bands of broadband and medium-band photometry are plotted over the PRISM  spectra, with broadbands shown in black and medium bands shown with colors corresponding to the those of the medium-band filter transmission curves at the top of the figure. The SED-fitted stellar mass, spectroscopic redshift, \haew, and Balmer break strength are listed for each galaxy at the left of each panel. The Balmer break and some notable emission lines are indicated. Bottom: Stacked Balmer breaks and H$\alpha$ emission for all 43 of the PRISM spectra in the sample. The spectra were locally normalized, interpolated to a common wavelength grid, and then combined. The spectra exhibit notable Balmer breaks; the median Balmer break strength of the sample is $\sim1.8$.}
    \label{fig:datasmall}
\end{figure*}

Here we present a broadband-selected sample of JWST PRISM spectra of low-mass galaxies ($7.5<\log M_\star/M_\odot < 9.5$) at cosmic noon ($1<z<3$). The sample was selected using photometry from the Ultradeep NIRSpec and NIRCam ObserVations before the Epoch of Reionization \citep[UNCOVER,][]{Bezanson2024} and MegaScience \citep{Suess2024} programs, which include imaging in all 20 medium- and broad-band filters of JWST/NIRCam over 30 arcmin$^2$ in Abell 2744. This sample provides a unique opportunity to study the dormant phase of low-mass, high-$z$ burst cycles. 

The paper is structured as follows. In Section~\ref{sec:data} we introduce the photometric catalog used to select followup sources and the main sample of PRISM spectroscopy. In Section~\ref{sec:sfhs}, we present the toy model for bursty star formation histories, use it to generate synthetic populations of galaxies, and discuss how the resulting observables depend on the burst parameters. In Section~\ref{sec:constraints}, we compare the generated distributions to the data, demonstrating that long timescale bursts are required to reproduce the observed distribution of Balmer breaks. Finally, we discuss the implications of our conclusion in the context of galaxy burstiness and physical frameworks of star-formation regulation in Section~\ref{sec:disc}.

Throughout this paper, we assume a standard $\Lambda$CDM WMAP9 cosmology \citep{Hinshaw2013} with $H_0 = 69.32$ km s$^{-1}$ Mpc$^{-1}$, $\Omega_m=0.2865$, and $\Omega_\Lambda=0.7135$. All magnitudes are reported in the AB system \citep{Oke1983}.

\section{Data} \label{sec:data}

\subsection{UNCOVER and MegaScience}\label{subsec:data_phot}
In this work, we use imaging, photometry, and spectroscopy from the JWST UNCOVER Treasury Program \citep{Bezanson2024} and the Medium Bands, Mega Science survey \citep[``MegaScience",][]{Suess2024}. UNCOVER obtained JWST/NIRCam broadband imaging of the Abell 2744 cluster in seven bands (NIRCam F115W, F150W, F200W, F277W, F356W, F410M, and F444W) and MegaScience imaged the same footprint in 13 additional bands (NIRCam F070W and F090W broadbands and F140M, F162M, F182M, F210M, F250M, F300M, F335M, F360M, F430M, F460M, and F480M medium bands) so that the combined catalog contains JWST imaging and photometry in all 20 NIRCam broad and medium bands. The UNCOVER and MegaScience catalogs make use of additional publicly available JWST data in Abell 2744 including imaging from ERS-GLASS \citep{Treu2022ApJ}, DD-2756 (PI: Chen), MAGNIF (GO-2883, PI: Sun), GO-3538 (PI: Iani), and ALT \citep[GO-2883,][]{Naidu2024}.

The reduction of the imaging for UNCOVER and MegaScience is detailed in \citet{Bezanson2024} and \citet{Suess2024}, respectively. The construction of the photometric catalog is described in \citet{Weaver2024}. The UNCOVER program also includes NIRSpec/PRISM spectroscopy selected based on the NIRCam photometry and imaging. The full prioritization scheme, mask design, observing, and reduction process for the UNCOVER PRISM sample is described in \citet{Price2025}, although the PRISM spectra used in this work were not included in the \citet{Price2025} release for reasons we discuss in the the following section.  

We use galaxy properties measured from SED fits to the photometry performed as presented in \citet{Wang2024}, which relies on the \texttt{Prospector-$\beta$} model \citep{Wang2023} and includes corrections  for lensing magnification \citep{Furtak2023}.  Specifically, values of stellar mass, star formation rate, metallicity, and attenuation for the galaxies in our sample are taken from the SED fits, which were performed with the spectroscopic redshifts fixed to those measured from the PRISM spectra. We note that the SED-fitting employs a non-parametric model for the SFHs with priors tied to the evolution of the cosmic star formation rate density.

\subsection{A Magnitude-selected Sample of PRISM Spectroscopy at Cosmic Noon}\label{subsec:data_spec}

We present a sample of broadband-magnitude selected (F200W $<27$ mag) PRISM spectra of low-mass galaxies ($7.5\lesssim\log M_\star/M_\odot \lesssim 9.5$) at cosmic noon ($1<z<3$). Our cleanly constructed sample represents a significant advance from what was possible before the launch of JWST. While nearly all previous low-mass spectroscopic samples at this epoch were emission-line selected and therefore highly incomplete, ours is based on rest-frame optical continuum emission and so includes numerous low-mass galaxies with low star formation rates (see \autoref{fig:samplesummary}).

While our sample includes sources with SFRs far below previous limits (compare our lower limit of $\log$SFR$_{10} \approx -2$ to that of $\log$SFR$_{\text{H}\alpha} \approx -0.5$ from \citet{Atek2022}), we are not fully complete to low-mass sources with exceedingly low SFRs, as we will discuss further in Sections~\ref{sec:constraints} and \ref{sec:disc}. In particular, our selection is not sensitive to quenched systems and in fact only one of the low-mass ($\log M_\star/M_\odot<9.5$) galaxies in our sample (\#21055) would be classified as quiescent by classical UVJ \citep{Whitaker2011} or NUVJr \citep{Ilbert2013} metrics. While quenched systems undoubtedly exist, evidence suggests that environment-driven low-mass quenching does not begin in earnest until lower redshifts \citep[$z\lesssim1.5$,][]{Cutler2025} and so old, quenched objects likely constitute a small portion of the overall population at this epoch.

Our uniform sample of filler spectra was made possible by technical issues that impacted two of the seven originally planned UNCOVER NIRSpec/PRISM visits over 2023 July 31 -- August 2. Repeat observations of the two affected masks were approved for 2024 July 30 -- 31, presenting an exciting opportunity to design a cleanly selected filler sample based on the 20-band photometry from UNCOVER and MegaScience and the resulting precise photometric redshifts. Sources with F200W~$<27$~mag and $1<z_\text{phot}<3$ were uniformly weighted as a filler sample and ultimately 57 were placed onto a mask and observed. 

While the \ha\ line falls into the F200W filter between $1.7 \lesssim z \lesssim 2.4$, the filter is broad and largely represents the optical continuum emission. There are 15 sources in this redshift window, but the \ha\ emission contributes only $\sim15\%$ of the flux on average, corresponding to an increase in F200W magnitude of $\sim0.15$ mags. All but one (\#36200) of these source have F200W magnitudes well above the 27 mag cutoff (F200W $<$ 26.7 mag) and also have F356W $<27$ mag, and so would have been included based on the current selection without the emission-line contribution and would have been included if the selection had relied on the uncontaminated F356W filter. The \ha\ contamination in the F200W filter does not bias our sample towards more heavily star forming sources.

For the purposes of our science, we only consider sources with $\log M_\star/M_\odot < 9.5$, excluding 13 of the 57 sources. As we discuss further below, we treat the sample of galaxies as a uniform ensemble; this assumption would not be reasonable if we included the high-mass objects, which are more metal- and dust-rich and are commonly affected by AGN activity. While there is uncertainty in the stellar mass estimates ($\sigma\approx0.13$), the sample would be nearly identical if we selected sources with F200W $>23$ mag. 

We note that our selection is slightly complicated by the magnification in the A2744 lensing region. The majority of our sources are minimally or moderately magnified ($\mu\lesssim2$), but we remove one source with extreme magnification ($\mu\approx11$) from our sample. The final sample consists of 43 low-mass cosmic-noon galaxies with PRISM spectra. The distribution of redshift, mass and specific star formation rate (sSFR) are shown in \autoref{fig:samplesummary}. The sources span a range of sSFRs and are representative of the photometric parent sample. A subset of the spectra, along with stacks of the sample in the Balmer break and \ha\ region, are shown in \autoref{fig:datasmall}. The full set of spectra is presented in \autoref{app:data}.

The PRISM spectra used in this work were accessed via the DAWN JWST archive (DJA)\footnote{\url{https://s3.amazonaws.com/msaexp-nirspec/extractions/nirspec_graded_v3.html}} and were reduced with \texttt{msaexp} \citep{Brammer2023}. The reduction and background subtraction strategies are described in \citet{Degraaff2025} and our data were reduced with a global background subtraction.

\subsection{Measuring \haew}

We measure \ha\ equivalent width (EW) directly from the PRISM spectra. Due to the low resolution, it is not possible to separate the \ha\ emission from the neighboring [\ion{N}{2}]$\lambda\lambda$6548, 6584 lines. Low-mass galaxies generally have lower gas-phase metallicities \citep{Tremonti2004}, so this contamination is not expected to be significant; the mass-metallicity relation presented in \citet{Sanders2021} predicts $12 + \log(\text{O/H})\approx 8.2$ for a galaxy with $\log(M_\star/M_\odot) = 9$ at $z=2.3$, corresponding to [\ion{N}{2}]$\lambda$6584/H$\alpha \approx 0.05$. Metallicity also has a second-order correlation with SFR and so this value should be even lower for the galaxies with the strongest H$\alpha$ EW. 

While it is technically possible to deblend the neighboring [\ion{S}{2}]$\lambda\lambda$6717,6732 lines, the coarse and wavelength-dependent resolution of the PRISM spectra make this challenging as it would require fitting two gaussian profiles to $\sim$10 wavelength elements, with variation in the fitting precision across redshift. Instead, as we describe below, we choose to integrate the \ha\ EW directly from the spectrum after fitting for the adjacent continuum. To illustrate the impact of [\ion{S}{2}] removal, we run a brief experiment, fitting a single gaussian profile and a double gaussian profile to the \ha\ region in each of the PRISM spectra. We find that deblending [\ion{S}{2}] decreases the EW by 0.5\%, a negligible effect as compared to the typical 10\% uncertainty.

The emission we measure is therefore the total emission of \ha\ and all adjacent lines within the specified wavelength range, uncorrected for stellar absorption.  We label these values as \haew\ for simplicity of notation, but caution that these values do not represent isolated \ha\ emission. We repeat an identical measurement procedure below on the synthetic spectrum, so this decision does not affect the data-model comparison.

For each spectrum, we estimate the continuum under \ha\ $(f_{\nu, \text{cont}})$ by fitting a power law to the adjacent regions of the spectrum: $0.56\,\mu m<\lambda_\text{rest}<0.64\,\mu m$ and  $0.69\,\mu m<\lambda_\text{rest}<0.76\,\mu m$. We perform the power law fit using the Markov Chain Monte Carlo package \texttt{emcee} \citep{Foreman-Mackey2013}. Next, we integrate the spectrum to measure the EW:
\begin{align}
    \text{H}\alpha\ \text{EW}_\text{obs} = \int_{\lambda_{0}}^{\lambda_1} 1-\dfrac{f_\nu(\lambda)}{f_{\nu, \text{cont}}(\lambda)} d\lambda
\end{align}
where $\lambda_0=(1+z)\cdot 0.62\,\mu m$ and $\lambda_1=(1+z)\cdot 0.71\,\mu m$. The rest-frame EW is then \haew\ $\equiv \text{H}\alpha\ \text{EW}_\text{rest} = \text{H}\alpha\ \text{EW}_\text{obs}/(1+z).$

\begin{figure}[t]
    \centering
\includegraphics[width=1\linewidth]{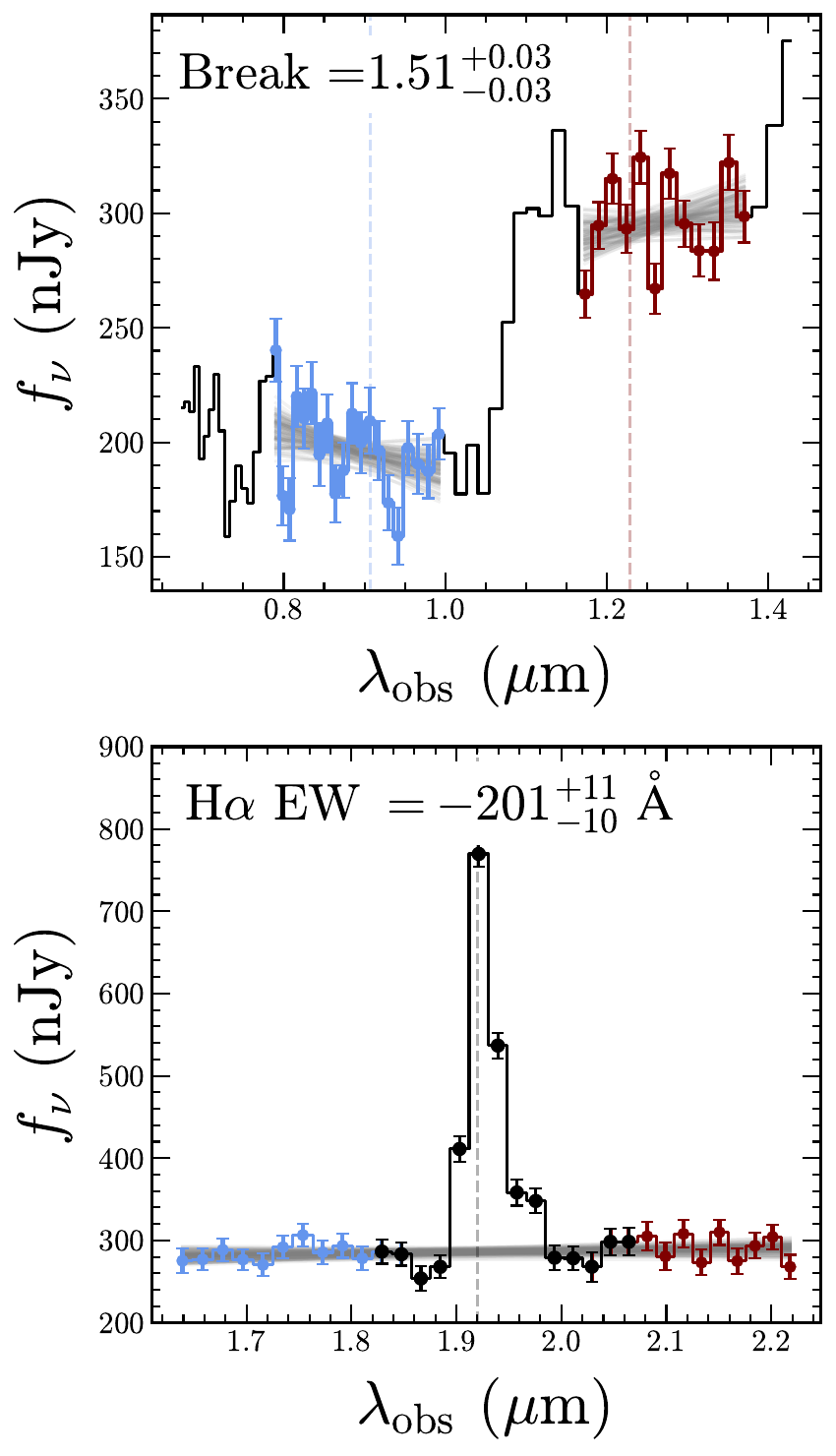}
    \caption{An example of the method we employ to measure Balmer break strength and \haew\ from the PRISM spectra. Left: The blue and red regions indicate the wavelength range over which the continua are fit on either side of the break. The vertical dashed lines show the wavelengths at which we evaluate the break strength. Right: The blue and red points indicate the wavelength range used to fit the continuum under \ha. The black points show the integration range we use to calculate the EW. }
    \label{fig:measurements}
\end{figure}

The 1$\sigma$ uncertainty on the EW is based on the posterior probability distribution, which is derived from the power law MCMC fit to the continuum. The continuum fit is the dominant source of uncertainty and the contribution from the uncertainty in the PRISM flux values over the integration range is negligible by comparison. An example of the \haew\ measurement is shown in \autoref{fig:measurements}.

\subsection{Measuring Break Strength}
To measure the strength of the Balmer break, we fit power laws to small, continuum-dominated regions on either side of the break and then take the ratio of the power laws evaluated at specific rest-frame wavelengths. We define the blue region as $0.27\,\mu m<\lambda_\text{rest}<0.34\,\mu m$ and the red region as $0.40\,\mu m<\lambda_\text{rest}<0.47\,\mu m$. We define the strength of the Balmer break as
$$\textrm{Balmer Break} = \frac{f_\nu(\lambda_\textrm{red, rest})}{f_\nu(\lambda_\text{blue, rest})}$$
with $\lambda_\text{blue, rest}= 0.31$\,$\mu$m and  $\lambda_\text{red, rest}= 0.42$\,$\mu$m. 

We take this power-law fitting approach instead of a simple ratio of median continuum values on either side of the break due to the sparse wavelength sampling of the PRISM spectra. A typical window of width 100\,\AA\ corresponds to two data points at PRISM resolution. Because of the wider wavelength windows, the spectral shape can influence the measurement along with the actual break strength, making this definition more susceptible to dust and metallicity than standard indices \citep[e.g.,][]{Balogh1999, Binggeli2019, Wang2024b}, which are poorly suited to the PRISM resolution of the Balmer break at these redshifts. However, our low-mass galaxies are minimally affected by dust attenuation (see further discussion in \autoref{app:dust}) and so the variation in the overall slope of the galaxy continua are less significant in our sample than for higher-mass or more strongly attenuated samples. While the continuum slope contributes somewhat to the break strength measured using our definition, we find our results presented below are robust to slight adjustments in the definition. We ultimately use the definition presented above as we find that using smaller wavelength ranges decreases the measurement precision and using wavelength ranges that are closer together decreases the stability of the measurement.

As for the \ha\ emission, the power-law fits are performed using \texttt{emcee} and the uncertainty in the break strength is derived from the posteriors of the fitted power law parameters.

\section{Modeling Bursty Populations} \label{sec:sfhs}

Though it is not feasible to recover each individual galaxy's star formation history from its spectrum \citep[see, for example][]{Wang2025}, the distributions of \haew\ and Balmer break strength we measure from the PRISM sample can provide constraints on the population-level characteristics of star formation in the observed galaxies. Previous works have derived such constraints by comparing UV and \ha\ emission \citep[e.g.,][]{Glazebrook1999, Weisz2012, Emami2019} and recent work has demonstrated the potential of expanding this analysis to include additional age-sensitive spectral features \citep{Iyer2024, Wang2025}. 

We adopt the assumption that star formation in all of the low-mass galaxies in our sample is governed by similar physical processes, which lead to duty-cycles on comparable timescales and with comparably sized deviations from the SFMS for the entire population. Using a simple model for periodically rising and falling star formation, we can then generate sets of synthetic spectroscopy for various choices of burst timescale and amplitude. Comparing the data with the distributions of \haew\ and Balmer break strength in these synthetic populations should reveal preferred and disallowed regions of the burst parameter space.

Below, we generate these simulated populations and measure their spectral features, investigating how their distributions depend on changing burst parameters.

\subsection{Bursty SFHs} \label{subsec:sfhs}

In order to construct a sample of simulated dwarf spectra and consequently simulated \haew\ and Balmer break distributions, we first develop a simple parametrization of bursty star formation histories. There are a number of important considerations, including the long-timescale rise of the SFHs with redshift and the shape of the bursts. The model we present is simplistic, but captures a galaxy's general SFR evolution and is neatly dependent on the timescale and amplitude of the bursts. These SFHs are not meant to be optimally realistic, but are instead designed to provide an interpretable explanation for the observed distribution of spectral properties in the sample. 

First, given an input stellar mass and redshift, we use the redshift-dependent SFMS presented in \citet{Popesso2023} to obtain the SFMS for such a galaxy as a function of time, $\text{SFMS}(M_{\star, z=z_\text{input}}, t)$. We choose to use this SFMS fit due to its convenient parameterization of the SFMS as a function of both mass and redshift, but note that at $1<z<3$ for $\log M_\star/M_\odot>8$ it is largely consistent with other SFMS fits \citep{Clarke2024}. A different choice of SFMS would impact only the long-term slope of the SFHs, which is expected to have a small impact on the derived Balmer break and \haew\ distributions, especially when compared to the dependence of these distributions on burst timescale and amplitude \citep{Wang2025}.

To add bursts to the slowly rising SFH, we first consider the shape and parametrization of a single burst. Each burst is modeled as a double-sided $\tau$-function--i.e. an exponential increase and an exponential decay, with each proportional to $e^{\pm t/\tau}$. The burst-duration and decay timescale are both determined by input parameter $\delta_\text{burst}$ in Myr such that a single burst cycle lasts for $\delta_\text{burst}$\ Myr and the decay timescale is fixed to $\tau=\delta_\text{burst}/20$. Disregarding the phase and amplitude of the bursts, the shape of the bursts with a given timescale $\delta_\text{burst}$ is given by
\begin{align}
    &\text{SFR}_\text{burst}(t, \delta_\text{burst}) =\\\nonumber
&\left\{
    \begin{array}{lr}
        e^{[t\ (\text{mod } \delta_\text{burst}) - \delta_\text{burst}/2]/\tau}& \text{if } t\ (\text{mod } \delta_\text{burst})< \frac{\delta_\text{burst}}{2} \\
        e^{-[t\ (\text{mod } \delta_\text{burst}) - \delta_\text{burst}/2]/\tau}& \text{if } t\ (\text{mod } \delta_\text{burst}) > \frac{\delta_\text{burst}}{2} \\
\end{array}
\right. .
\end{align}

The amplitude and lower SFR limit of the bursts are determined both by the general evolution of the galaxy on the SFMS and by the parameter \sigsf, the logarithmic deviation of the SFH above and below the main sequence, given in dex. Ultimately, we normalize the SFHs such that the total formed mass equals the input stellar mass (additional details are provided below), but before normalization, the burst maximum reaches \sigsf\ above $\text{SFMS}(M_{\star, z=z_\text{input}}, t)$ and the minimum reaches the same distance below. Finally, the phase of the bursts is determined by $t_0 = \delta_\text{burst} \cdot \frac{\phi}{2\pi}$ where $\phi\in[0,2\pi]$ such that $t_0\in[0,\, $\db]. The full parametrization is then determined by input parameters ($M_{\star, z=z_\text{input}}, \delta_\text{burst},A_\text{SFMS}, t_0$) such that
\begin{align}
    &\text{SFR}(t) = \text{SFMS}(M_{\star, z=z_\text{input}}, t)\times \nonumber\\
    &\left[\text{SFR}_\text{burst}(t-t_0, \delta_\text{burst})(10^{A_\text{SFMS}} - 10^{-A_\text{SFMS}}) + 10^{-A_\text{SFMS}}) \right].
\end{align}

Finally, the SFHs are normalized so that the total formed stellar mass is equal to the input stellar mass. We renormalize to the total formed stellar mass instead of the surviving stellar mass so that the SFHs are independent of the chosen stellar evolution model. The difference between the surviving stellar mass and total formed stellar mass in our modeled galaxies is small ($\sim0.1$ dex) and the spectral features we consider are continuum-independent, so this choice does not impact our results. The renormalization shifts the SFHs such that they are not always logarithmically symmetric about the SFMS. 

We caution that the parameter \sigsf, the logarithmic deviation of the maximum and minimum SFR from the SFMS (before normalization),  should not be equated with the often-discussed main-sequence scatter $\sigma$. Relating \sigsf\ to $\sigma$ is non-trivial and made less meaningful by our simplifying assumptions. In particular, the derived $\sigma$ would be highly tied to the shape of our burst model, which we have intentionally constructed to be more illustrative than maximally physically realistic. Perhaps more importantly, $\sigma$ would be strongly impacted by our treatment of the population as a unified ensemble with a single shared value of \db\ and \sigsf\ (detailed further in the following section). As we discuss in Section~\ref{sec:disc}, future work employing larger samples and less highly-parameterized burst models will be better suited to explore implied constraints on $\sigma$.

\begin{figure*}[th]
    \centering
    \includegraphics[width=1\linewidth]{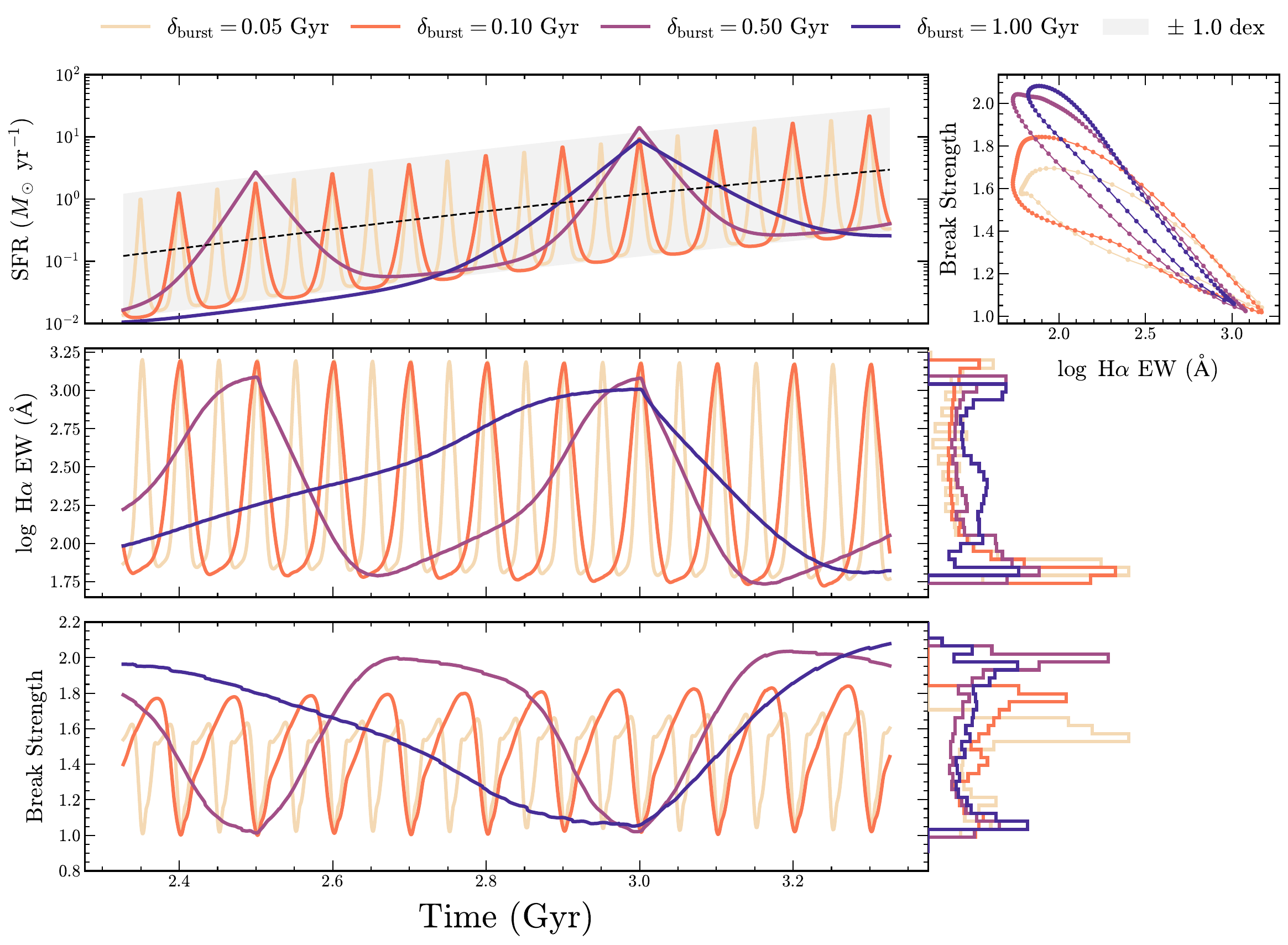}
    \caption{Bursty SFHs for a galaxy with $M_\star=10^9\,M_\odot$ at $z=2$ with fixed amplitude (\sigsf\ $=1$ dex) and varying timescale (\db). Top: The SFR over the most recent 1 Gyr for bursty SFHs with \db $=0.05, 0.1, 0.5,$ and 1 Gyr plotted in different colors. The dashed line shows the evolution of a burst-free galaxy on the \citet{Popesso2023} star-forming main sequence and the shaded region indicates the area 1 dex above and below the main sequence. Middle: The log \haew\ as a function of time for the various timescales with marginal distributions shown at the right. Bottom: The same as the middle panel, but for the Balmer break strength. Top right: The evolution of a galaxy following the various bursty SFHs in the Balmer break vs. log \haew\ space. The points shown are evenly time spaced and sampled over a full period in phase, with $t_0$ running from 0 to \db\ as described in Section~\ref{subsec:sfhs}. Strong breaks are only achieved in SFHs with long-timescale bursts; SFHs with shorter bursts always sustain a population of young, massive stars and so never develop strong Balmer breaks, which are characteristic of a stellar population dominated by A stars.}
    \label{fig:sfhs_db}
\end{figure*}

\begin{figure*}[th]
    \centering
    \includegraphics[width=1\linewidth]{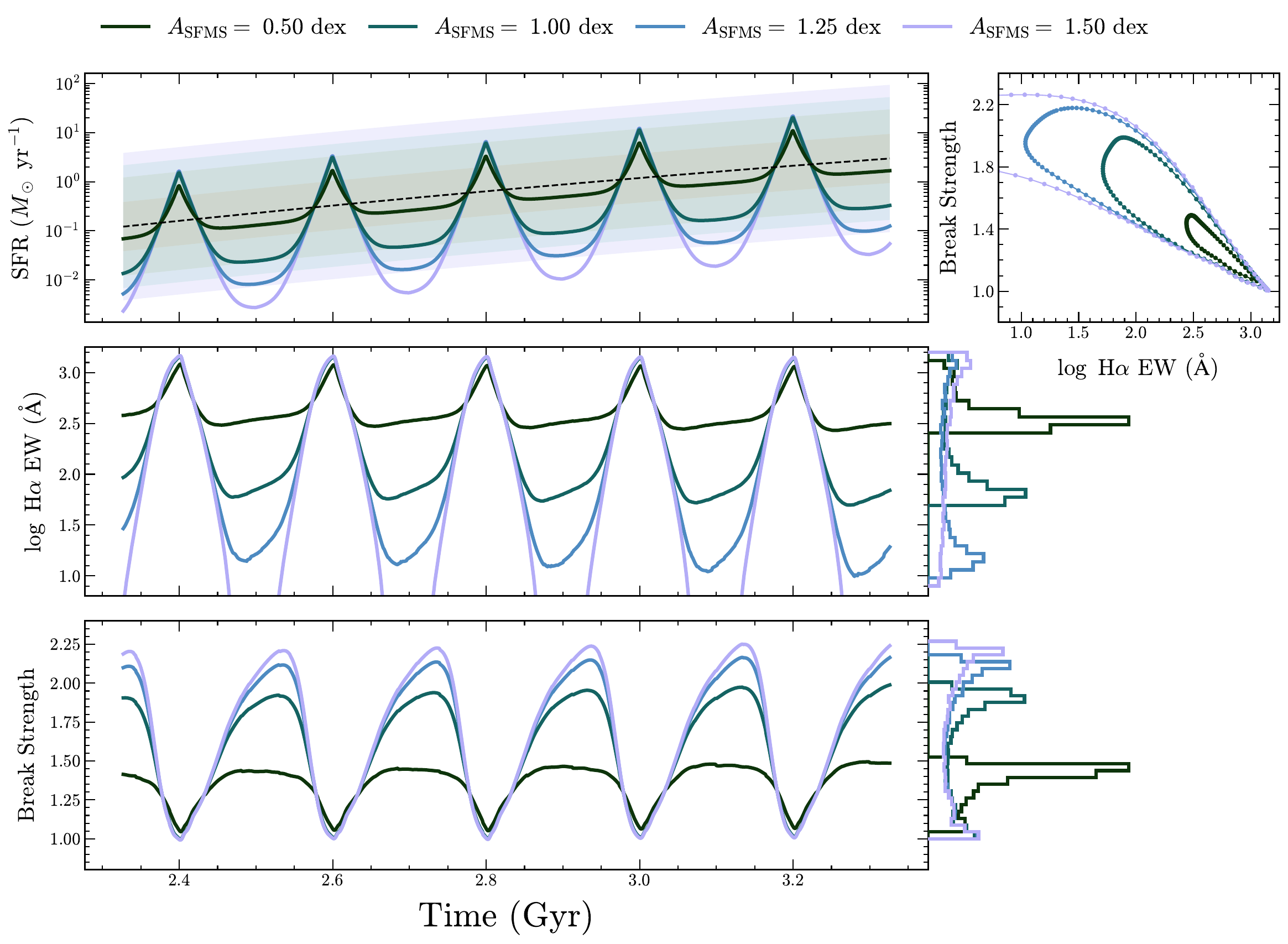}
    \caption{The same as \autoref{fig:sfhs_db}, but varying \sigsf\  with \db\ held constant at 200 Myr. The shaded regions show the symmetric logarithmic deviations of \sigsf\ from the main sequence, which are not matched to the range of the final SFHs due to the mass normalization. Strong breaks are only achieved by SFHs with significant deviation from the SFMS. Low-amplitude bursty SFHs never reach sufficiently low SFRs, even in their dormant periods.}
    \label{fig:sfhs_sig}
\end{figure*}

Examples of SFHs for a galaxy with $\log(M_\star/M_\odot)=9$ at $z=2$ with various values of $\delta_\text{burst}$ at fixed \sigsf\  are shown in \autoref{fig:sfhs_db} and for various values of \sigsf\  at fixed $\delta_\text{burst}$ in \autoref{fig:sfhs_sig}.

Although the toy SFHs are tied to the SED-derived stellar masses, which are fit with non-parametric SFHs, we note that subsequent analysis does not depend strongly on the stellar mass. The spectral features we rely on are both continuum normalized and therefore largely insensitive to the absolute luminosity or stellar mass of the systems. Although the parametric SFHs assume a continuity prior \citep{Leja2019}, previous work has shown that SED-derived stellar masses are not sensitive to the assumed burstiness of the SFHs \citep{Wang2024_prospector}. Regardless of the SFH-shape prior, the non-parametric SFHs used in the SED-fitting are much more coarsely sampled than the models we present (at most 5 bins over a Gyr of lookback time) and therefore cannot capture the presence or lack of the bursts we describe. We also note that the long timescale evolution of the SFHs are consistent with our models, as the priors used in the SED-fits are based on the evolution of the cosmic star formation rate density, akin to our choice to tie the SFH models to the evolving SFMS. We compare the SED-derived star formation rates to the our modeled SFRs in Section~\ref{subsec:sfms}.

\subsection{Generating Synthetic Populations}\label{subsec:modelgen}

To determine whether the Balmer break--\haew\ distribution of our sample can constrain the timescale or amplitude of SF bursts on a population level, we generate synthetic ensembles of representative galaxies with uniformly bursty SFHs. For a given choice of timescale ($\delta_\text{burst}$) and amplitude (\sigsf), we construct bursty SFHs as described above for each $(M_\star,z)$ pair from our spectroscopic sample, using a temporal resolution of 1 Myr. We vary the phase ($t_0$) for each galaxy's SFH, using 50 values of $t_0$ per galaxy with $t_0$ running from 0 to $\delta_\text{burst}$ so that for each set of $\delta_\text{burst}$ and \sigsf, we create an ensemble of $50\cdot N_\text{gal}$ bursty SFHs. 

We use the Python implementation of the Flexible Population Synthesis Code \citep[FSPS,][]{Conroy2009, Conroy2010, fsps} to generate a synthetic spectrum for each SFH. We use the MILES stellar library, MIST isochrones, and a \citet{Chabrier2003} IMF. Nebular emission is modeled using the photoionization code \texttt{CLOUDY} \citep{Ferland2013} as described in \citet{Byler2017}. The gas-phase and stellar metallicity are set to be equal, and are randomly drawn from the stellar metallicities measured from the SED fits to the photometry of the galaxies in the PRISM sample (as described in Section~\ref{subsec:data_phot}). The metallicities range from $\log Z/Z_\odot=-1.79$ to $-0.21$ with a median of $\log Z/Z_\odot = -0.97$. Attenuation is added using a \citet{Kriek2013} attenuation curve, which applies attenuation to all stars and additional attenuation to young stars. We fix the relative attenuation – i.e. setting $\tau_1 =\tau_2$. As with the metallicity, for each simulated spectrm, $\tau$ is randomly drawn from the distribution of $\tau$ from the SED fits. The attenuation ranges from $A_V$ = 0.01 to 0.88 mag, with a median value of $A_V=0.21$ mag. We discuss the robustness of our results to the choice of metallicity and attenuation in \autoref{app:dust}. In general, we find our results are minimally affected by reasonable adjustments to these decisions. 

To measure \haew\ and Balmer break strength from the synthetic spectra, we smooth the spectra to the JWST/PRISM resolution by convolving with the instrumental line spread function\footnote{\url{https://jwst-docs.stsci.edu/}}. We then measure the \ha\ emission and Balmer break in the same way we measure these values from the observed spectroscopy as described in Section~\ref{sec:data}. However, instead of fitting the continuum power laws with \texttt{emcee}, we instead perform a simple least-squares minimization, which is less computationally expensive.

\subsection{Dependence of Distribution of Balmer Break Strength and \haew\ on Burst Parameters}\label{subsec:model_distributions}
    
Before comparing the model distributions with the measured values from our spectroscopic sample, we first explore how the resulting \haew\ and Balmer break depend on the burst parameters described above. Below we show that significant and extended deviations below the SFMS are required to develop strong Balmer breaks. 

\subsubsection{Effect of Varying \db}

First, we fix \sigsf$=1$ dex and vary \db\ for a galaxy with $M_\star = 10^9\,M_\odot$ at $z=2$ as shown in the top panel of \autoref{fig:sfhs_db}. We generate synthetic spectra as described above, but fix the metallicity to 10\%\,$Z_\odot$ and apply no attenuation to illustrate the direct impact of SFH variations on the spectral features.

\textit{\haew:} For longer burst timescales, the \haew\ peaks at lower values as shown in the middle panel of \autoref{fig:sfhs_db}. While the \haew\ roughly traces a galaxy's sSFR, the peak of which does not vary significantly among the SFHs, it is also dependent on the continuum, which is elevated for the long timescale bursts that rise slowly and build a sustained young stellar population. The precise shape of the \haew\ evolution -- and the marginals -- are tied to the exponential increase and decay of the bursts. While the general trends would be similar for different burst shapes (for example, a top hat function or an exponential decay following a discontinuous jump), the shape of the curve and one-dimensional distributions would be affected.

\textit{Balmer break strength:} The bottom panel of \autoref{fig:sfhs_db} demonstrates the power of the Balmer break to distinguish among potential burst timescales. A galaxy with star formation dominated by short bursts (50--100 Myr) never develops a significant Balmer break ($\gtrsim 1.8$). While the galaxy is dormant, the Balmer break strength increases as the massive stars die off and the A-star population begins to dominate the emission. But before the Balmer break is able to develop further (to reach a strength of $\sim2$), the galaxy reignites star formation and the light-weighted age falls again. The peak Balmer break strength is higher for SFHs with longer bursts, but the maximum Balmer break strength does not increase continuously with \db. Once the least massive B stars have left the main sequence, which takes on the order of 100--500 Myr, the A stars dominate the emitted light and the Balmer break reaches its maximum strength. The lowest-mass A stars will survive for up to an additional 1--2 Gyr and so there will be little change in the Balmer break over that interval. The distribution of Balmer break strength for the SFHs with \db\ of 500 Myr and 1 Gyr therefore peak at similar values, as expected. The main difference is in the shape of the distribution, again owing to the slow rise and fall of the SFR in the long timescale case. 

\textit{\haew\ and Balmer break strength:} 
In addition to the one-dimensional differences described above, it is clear from the top right panel of \autoref{fig:sfhs_db} that the slope and spread of the \haew--break strength distribution are dependent on the burst timescale. There is significant overlap among the various \db\ tracks at high \haew, but at low \haew, the distributions occupy distinct regions of the parameter space. Given that the galaxies spend the most time at low \haew\ (see the density of evenly time-spaced points), this distribution offers a promising opportunity to use a magnitude-limited sample to distinguish among the potential timescales. 

While true SFHs are undoubtedly more complex than our toy model, the dependence of the Balmer break-strength distribution (its peak and shape) on $\delta_\text{burst}$ is revealing. Most simply, if a population of galaxies contains a significant number of strong Balmer breaks, it is highly unlikely that star formation in the population is dominated by burst-free star formation (\db$\rightarrow0$) or star formation with short timescale bursts.

\begin{figure*}
    \centering
    \includegraphics[width=1\linewidth]{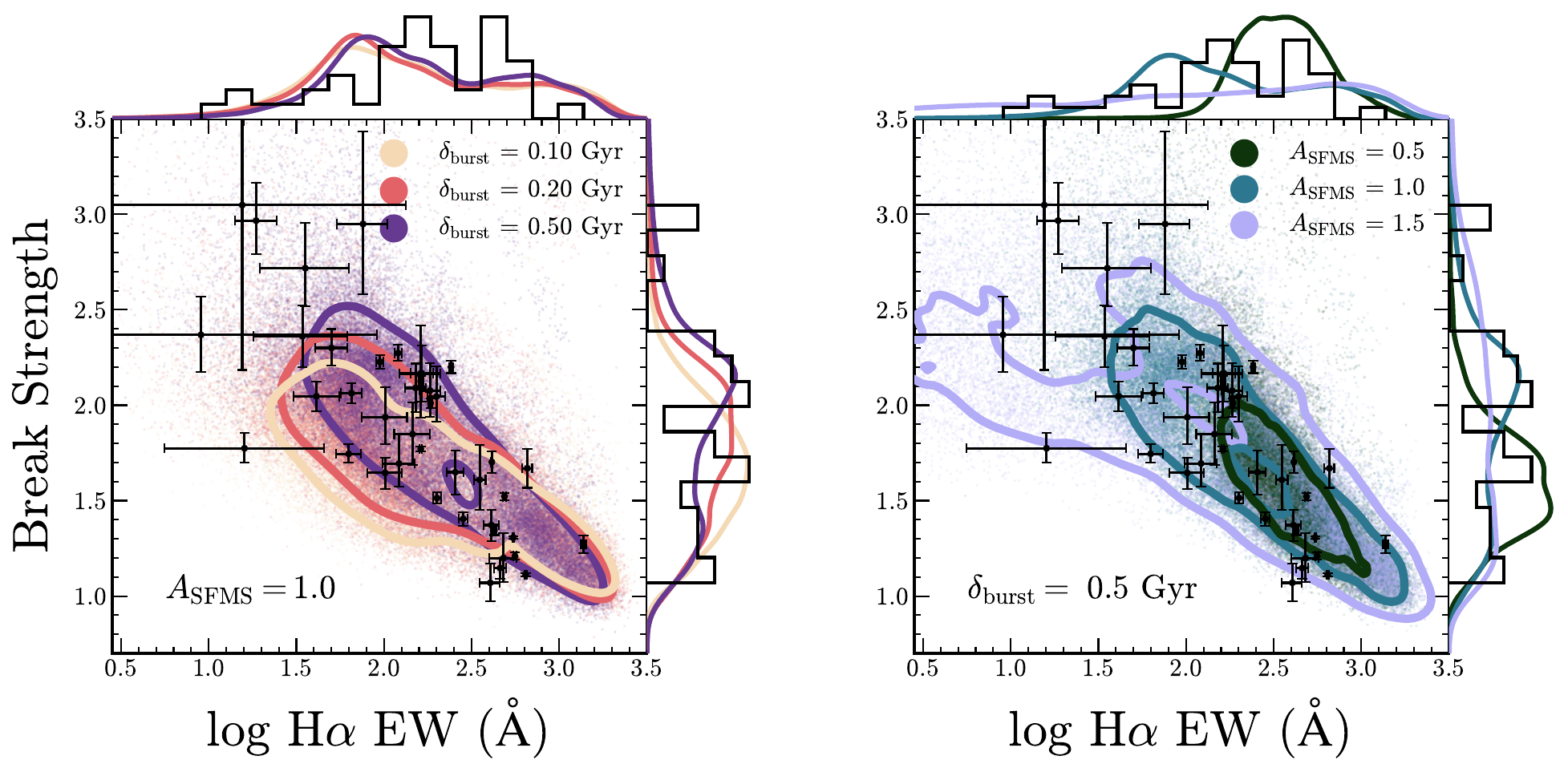}
    \caption{The distribution of \haew\ and Balmer break strength measured from our magnitude-limited spectroscopic sample implies that star formation in this low-mass cosmic noon population is likely dominated by bursts with long timescales and significant deviations below the star-forming main sequence. Left: The black points show the \haew\ and Balmer break strength we measure from the PRISM spectra as described in Section~\ref{sec:data}. The points and 1$\sigma$ contours show the distributions of the simulated galaxy populations with SFHs with \sigsf$=1.0$ dex and \db\ of 100 Myr, 200 Myr, and 500 Myr with jitter applied to match the uncertainty in the PRISM measurements. Marginal distributions are shown for the break strength on the right and for the \haew\ on the top of the panel. While the \haew\ distributions of the simulated populations are similar, the distributions of the Balmer break are clearly inconsistent with the data for \db $<$ 200 Myr. Right: The same as the left panel, but for simulated populations with SFHs with \db$=$ 500 Myr and \sigsf\ of 0.5, 1.0, and 1.5 dex. The distribution for \sigsf$=0.5$ dex is clearly inconsistent with the data. \sigsf$=1.0$ dex matches the data well and \sigsf=$1.5$ dex includes more low \haew\ systems than we observe, but we cannot say conclusively whether this is reflective of our incompleteness to such systems.}
    \label{fig:2d-dist}
\end{figure*}

\subsubsection{Effect of Varying \sigsf}

Next, we explore how the \haew\ and Balmer break are affected by changing the burst amplitudes for a fixed burst timescale. In \autoref{fig:sfhs_sig}, we fix  $\delta_\text{burst}=$ 200 Myr and vary \sigsf\ for a galaxy with $M_\star=10^9\,M_\odot$ at $z=2$. Due to the renormalization of the SFHs to recover the input stellar mass, the SFHs are not logarithmically symmetric about the main sequence. Above \sigsf\  $=0.5$, all of the SFHs reach similar maximum star formation rates and the main distinction among the various \sigsf\  values is the minimum SFR reached during the dormant period of the cycle.

\textit{\haew:} These differences in star formation range are reflected in the evolution of the \haew, which peaks at a similar value for \sigsf\  $\ge1$. The minimum \haew\ declines with increasing \sigsf\, with SFHs with \sigsf\ $\ge1$ reaching minima below $\sim 80$\,\AA. The SFH with \sigsf $=1.5$ dex yields spectra with \ha\ absorption in its most dormant phases while the SFH with very small deviations from the SFMS (\sigsf\ $=0.5$ dex) never reaches \haew\ below $\sim300$\,\AA. 

\textit{Balmer  break strength:} The low-amplitude bursty SFH also does not allow for the development of significant breaks. In this case, the galaxy's ``dormant" period is still quite active and so it sustains a significant population of young massive stars at all times and maintains a young light-weighted age and therefore a weak Balmer break. As \sigsf\  increases, the maximum Balmer break also increases, with the rate of increase plateauing above \sigsf\ $\approx1.25$ dex. 

\textit{\haew\ and Balmer break strength:} In the two-dimensional \haew--break strength distribution shown in the top right panel, we see that increasing \sigsf\  increases the range of parameter space covered in one cycle. As the SFHs approach low values of \sigsf\  (i.e. burst-free star formation), the distribution is restricted to a small region, with the \haew\ remaining high and the break strength low. Exceedingly high values of \sigsf\  (\sigsf\  $\gtrsim$ 1.5) lead to H$\alpha$ absorption during the dormant period of the cycle. Again, we see significant degeneracy among the \sigsf\  values in the high \haew\ region of the distribution. A sample of \ha-flux selected galaxies would not allow us to constrain the burst amplitude from this distribution.

As with the timescale variation, a clear conclusion emerges from this analysis, regardless of the simplistic nature of the models: significant deviations ($\gtrsim1$ dex) from the SFMS are required to develop strong Balmer breaks. It is therefore highly unlikely that a population of galaxies which contains strong Balmer breaks is dominated by constant star formation along the SFMS or bursty star formation with small drops below the SFMS. 

\section{Constraints on Burst Parameters from Data}\label{sec:constraints}

Having established a general intuition for the dependence of \haew\ and Balmer break strength on the burst parameters \db\ and \sigsf, we can compare the simulated population distributions generated in Section~\ref{subsec:modelgen} to the distributions measured from our spectroscopic sample as described in Section~\ref{sec:data}. While the models and the mode of comparison are both simplistic, the striking prevalence of strong Balmer breaks in our low-mass cosmic noon sample of PRISM spectra provides clear constraints on the star-formation behavior of the ensemble. In Section~\ref{sec:disc}, we discuss future extensions of this work that will include significantly larger samples and more rigorous statistical methods, but we emphasize here that even an initial analysis of the current sample reveals unambiguous conclusions about the nature of bursts in this population.

In the following sections, we treat the sample of galaxies as an ensemble. We assume that all of the galaxies follow the bursty SFH models defined in Section~\ref{sec:sfhs} and that they share a timescale \db\ and amplitude \sigsf. Though the assumed uniformity of the sources in \db\ and \sigsf\ is almost certainly overly simplistic, the framework of dynamic movement about the MS is well justified.  While it is possible that some of the galaxies we observe below the main sequence with strong Balmer breaks are permanently quenching, it is unlikely to be the case for more than a few sources. Over 40\% of the sources have Balmer break strength $\gtrsim2$ and a similar number have sSFRs below the median of the parent sample (see \autoref{fig:samplesummary}). Considering that permanent quenching is believed to occur rapidly, observing such a significant portion of the population in a brief and transient phase is highly improbable. We consider this and potential environmental effects in more detail in Section~\ref{subsec:environment}.

\subsection{Short Timescales and Low Amplitudes Excluded by \haew--Balmer Break Strength Distribution}

In \autoref{fig:2d-dist}, we compare the distributions of \haew\ and Balmer break strength measured from the PRISM sample to the distributions from the synthetic populations with various values of \db\ and \sigsf. In the left panel, we fix \sigsf$=1.0$ dex and show three choices of \db. The distributions for bursty SFHs with short burst timescales (\db$\lesssim$ 100 Myr) are not well matched to the data (a statistical comparison is discussed in the following section). The slope of the distribution is shallower and strong break strengths are never reached. As discussed in Section~\ref{subsec:model_distributions}, the range and shape of the \haew\ marginal distribution does not vary significantly with the burst timescale and all of the marginal distributions shown in \autoref{fig:2d-dist} agree reasonably well with the data. The shape of the \haew\ distributions are particularly dependent on the choice of burst shape for the SFH parametrization. For example, SFHs with bursts described by a single declining $\tau$ model with a sharp, discontinuous jump and exponential decay yield \haew\ distributions that appear less bimodal than those shown in \autoref{fig:2d-dist}. While the same is true for the Balmer break distribution, the \haew\ is much more sensitive to these changes. The strength of simulated \ha\ emission is also strongly model-dependent and based on unsettled science of massive-star ionizing-photon production \citep{Tacchella2022}. While the modeled Balmer break distributions are minimally affected by the choice of isochrone (i.e. Padova \citep{Girardi2000, Marigo2007, Marigo2008} versus MIST), the \haew\ distributions are changed significantly \citep[see also][]{Byler2017}. 

Our specific calculation of the \haew\ also includes adjacent blended lines and is therefore additionally reliant on metallicity. The Balmer break is more robust to many of these issues and so represents a promising path for constraining burstiness when it is measurable. In fact, \autoref{fig:2d-dist} shows that the Balmer break distribution does vary significantly with \db\ and only the longer timescale distributions shown match the data well. 

In the right panel of \autoref{fig:2d-dist}, we fix \db $=$ 500 Myr and vary \sigsf. The range of parameter space covered by the PRISM sample clearly excludes the population with \sigsf$=0.5$ dex. These small-amplitude bursts result in SFHs that are nearly constant and have no significant deviations below the main sequence; the galaxies in this synthetic population always maintain strong \haew\ and weak Balmer breaks. The highest value of \sigsf$=1.5$ is a better match to the Balmer break distribution of the data, but predicts a population of lower \haew\ that we do not observe. 

Under our assumption that star formation in the population of low-mass galaxies at cosmic noon is dominated by bursts with a single timescale and (mass-dependent) amplitude, our magnitude-limited sample of PRISM spectra clearly disfavors large regions of the available burst-parameter space. Most significantly, the distribution of Balmer break strengths that we observe indicates that star formation in these galaxies must be dominated by long timescale bursts with significant (but not exceedingly large) deviations below the main sequence.

\subsection{Constraints from Balmer Break Alone}\label{subsec:hists}

\begin{figure}
    \centering
\includegraphics[width=1.\linewidth]{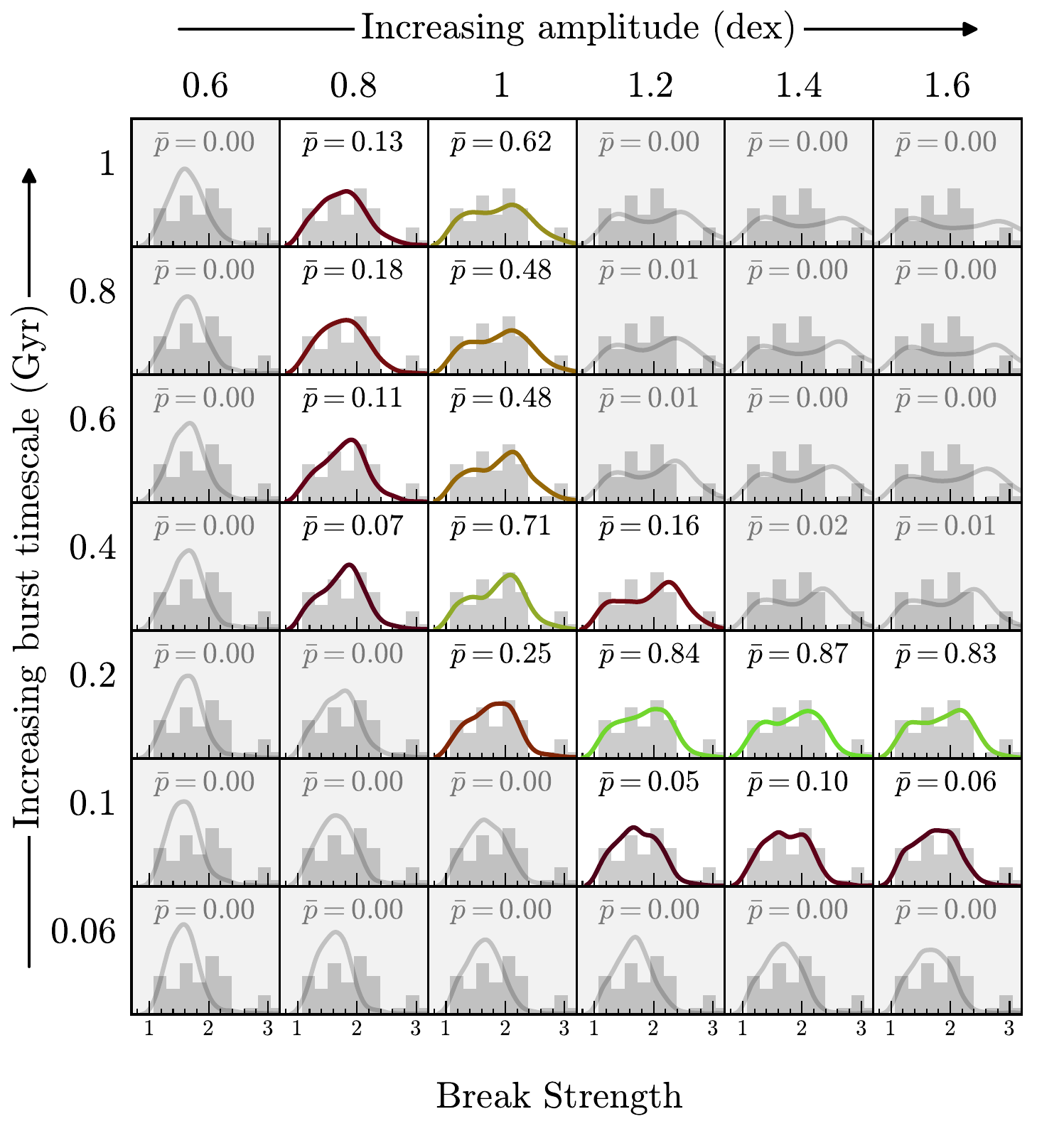}
    \caption{Distributions of Balmer break strength of simulated galaxy populations generated with various combinations of burst timescale \db\ and amplitude \sigsf. Each panel shows the simulated density distribution for a given \db\ and \sigsf\ as a solid line plotted over the histogram of the Balmer break strengths measured from the PRISM spectra. The value of \db\ increases with each row and the value of \sigsf\ increases with each column. To account for measurement uncertainty, we perturb the Balmer break strengths from the data around their best-fit values and perform a Kolmogorov-Smirnov test for each of the 1000 realizations. Each panel shows the median p-value from the 1000 KS tests and panels with simulated distributions that are inconsistent with the data ($p<0.05$) are shaded gray. Distributions with $p>0.05$ are colored by their p-value, with green distributions corresponding to higher p-values. The measured distribution of Balmer breaks appears to rule out bursty SF with short timescales and small deviations from the SFMS. }
    \label{fig:bb_hist}
\end{figure}

In the above section and in \autoref{fig:2d-dist}, we show only a small selection of burst-parameter combinations. To more thoroughly demonstrate the regions of \db-\sigsf\ parameter space that are allowed and excluded by our sample, we compare one-dimensional distributions of the Balmer break strength from the simulated ensembles to our observed population in \autoref{fig:bb_hist}. To evaluate the agreement between the observed and modeled distributions, we perform a Kolmogorov--Smirnov test on the Balmer break distributions. To account for uncertainty in the Balmer break measurement (particularly in the strong Balmer breaks as seen in \autoref{fig:2d-dist}), we perturb the measured Balmer breaks around their best-fit values by drawing from a normal distribution with $\sigma$ equal to the measurement uncertainty. For each realization, we measure the KS-test p-value comparing the perturbed data distribution to the predicted model distribution for each \db--\sigsf\ pair. We measure the median p-value from the 1000 realization and consider model distributions with $\bar{p}<0.05$ to be poorly matched to the data. The median p-values are shown in \autoref{fig:bb_hist}.

The distributions of Balmer break strengths shown in \autoref{fig:bb_hist} clearly demonstrate that star formation in the observed galaxies is not dominated by short bursts or bursts with small amplitudes. The simulated ensembles in both of these cases yield Balmer break strength distributions that are limited to values below $\sim$2, while the observed distribution reaches significantly higher values. For exceedingly high values of \db\ and \sigsf, the simulated Balmer break distributions are skewed to higher values than those of the observed galaxies. 

The observed Balmer break distribution clearly favors a region of parameter space with \db\ $\gtrsim$ 100 Myr and \sigsf\ $\gtrsim 0.8$ dex. We defer a more rigorous parameter optimization for future work, which will involve a significantly larger sample and more sophisticated generation of model SFHs. A larger dataset will allow us to better sample the underlying distribution and more confidently analyze its shape while a more flexible approach to generating SFHs will remove the dependence of the simulated distribution shape on the precise parameterization of the bursty SFHs.

We note that the reported results are driven by the large number of galaxies in the sample with strong Balmer breaks and they do not depend on a small number of galaxies with anomalous values; nearly half of the galaxies in the sample have Balmer break strengths greater than 2. We further demonstrate the robustness of our results to resampling in \autoref{app:bootstrap}.

\subsection{Considering Environment}\label{subsec:environment}
 
Galaxies that are interacting with close companions or undergoing mergers would violate our assumption that all of the galaxies in our sample follow a common mode of bursty star formation. In particular, if a significant number of the galaxies are in the process of environment-driven quenching, our interpretation of the Balmer break strength as indicating long burst timescales would be invalid. To investigate this potential complication, we first attempt to directly identify whether any of the galaxies in our sample have closely located neighbors with which they may be interacting. We use the publicly available catalog of spectroscopic redshifts from the JWST/NIRCam grism survey All the Little Things \citep[ALT,][]{Naidu2024}  to search for close neighbors and as an extreme upper limit on the environment, we also search for close companions in the full photometric redshift catalog from MegaScience. 

Only four of the galaxies in our sample have a close neighbor in the ALT catalog – defined as having $\Delta z/(1+z) < 0.005$ and being within 3\arcsec. These four galaxies are not clustered toward the high break, low \haew\ region of the distribution and in fact have somewhat low break strengths and high \haew. The 3\arcsec\ radius was chosen as in \citet{Matthee2024}, by identifying the minimum in the pair count distribution for sources in ALT with $1<z<3$. Increasing the radius beyond 5\arcsec\ yields additional galaxies with close neighbors, but even with a large matching radius of 20\arcsec, we find only 10 galaxies with pairs. As before, these galaxies are randomly distributed in the \haew-Balmer break space, with a slight clustering towards higher \haew. 

Using the MegaScience photo-z catalog and loosening the redshift requirement to $\Delta z/(1+z) < 0.02$ accordingly, we find that 11 of our galaxies have another source within 3\arcsec. However, only four of those close neighbors are more massive than the primary galaxy in our sample. And as with the ALT close companions, the galaxies with nearby neighbors in the MegaScience catalog have \haew\ and Balmer breaks that are distributed similarly to the sample as a whole, and tend to cluster in the high H$\alpha$, low break strength region of the parameter space. The results of this exploration indicate that environmental quenching is likely not significant for galaxies in our sample and further justifies our treatment of the sample as a coherent ensemble.

 In addition to our brief investigation of environment, existing evidence from the literature supports our conclusion that environmental effects on star formation are minimal at cosmic noon and the picture of low-mass environmental quenching in the local Universe \citep{Geha2012} cannot be extended to higher redshifts. \citet{Pan2025} identify a proto-cluster in the MegaScience photometric catalog at $z\approx2.6$ and find no discernible difference in the SFHs of the proto-cluster members versus the field sources. They conclude that at cosmic noon, because massive clusters have yet to virialize, the prevalence of environmental quenching is small compared to the nearby Universe. To assess the impact of the proto-cluster on our results, we repeat the above analysis with sources with $2.4<z<2.7$ removed and find the results are unchanged. Other recent studies have found little or no difference in the quenched fraction of low-mass galaxies ($9 \lesssim\log M_\star/M_\odot \lesssim 9.5$) in the field and in proto-clusters at $z\approx 1-3$, concluding that environmental effects have yet to dominate quenching \citep{Forrest2024, Edward2024}.

\section{Discussion}\label{sec:disc}
\subsection{JWST Enables Population-level Analysis of Dwarfs at Cosmic Noon}

\begin{figure*}
    \centering
\includegraphics[width=1\linewidth]{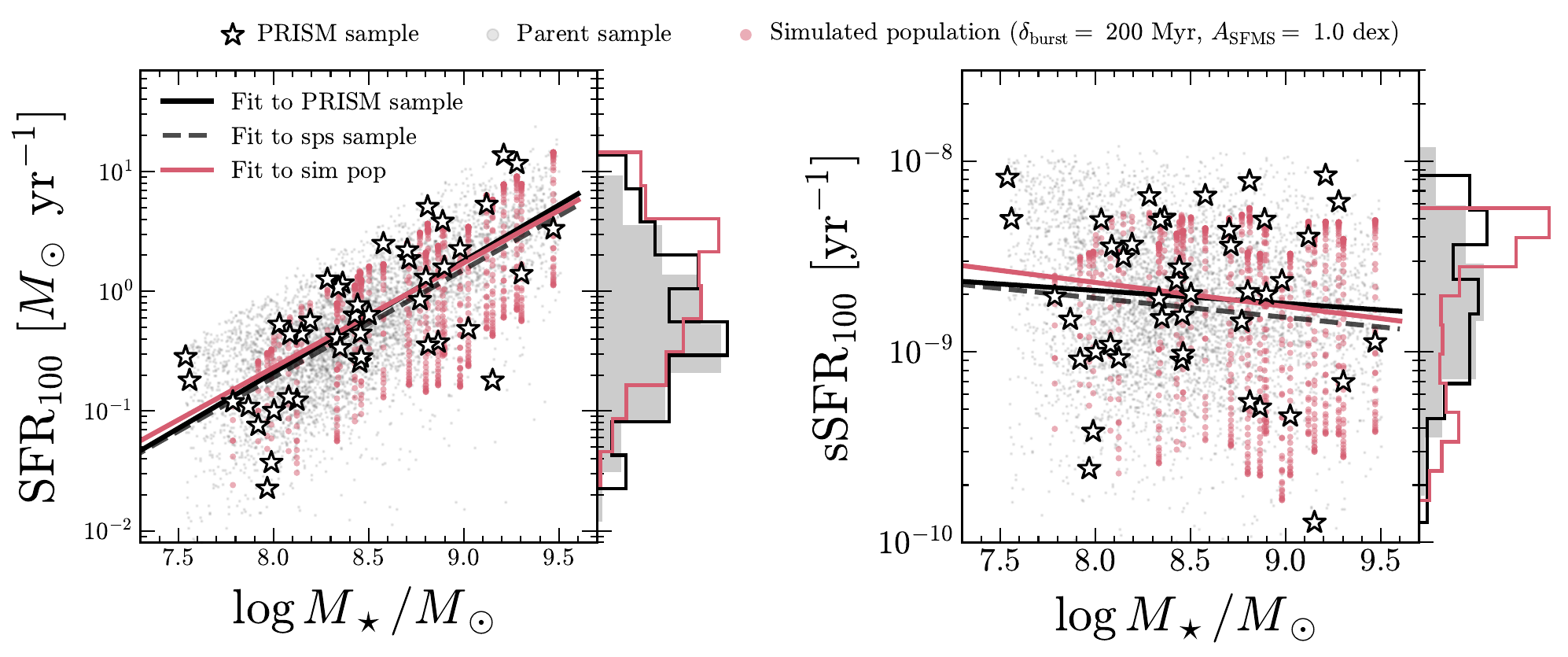}
    \caption{Left: Star formation rate versus stellar mass. The stars show the sources in the PRISM sample (excluding those with inaccurate photometric redshifts -- see Section~\ref{subsec:data_phot}) and the gray points show the sources in the parent sample with SFRs measured from SED fits to the 20-band MegaScience photometry. The red points show the SFRs averaged over the most recent 100 Myrs for the model bursty SFHs with \db\ = 200 Myr and \sigsf\ = 1 dex. Only models that yield synthetic SEDs with F200W $<$ 27 are shown. The lines show the SFMS fit to the three samples, which are largely in agreement. Right: The same as the left panel, but for the specific star formation rate. Both the PRISM sample and the parent sample contain significant populations well below the SFMS. The models, while less representative in the bursting phase above the SFMS, capture the dormant phase well.}
    \label{fig:sfms}
\end{figure*}

Before JWST, broadband-magnitude-limited spectroscopic samples of dwarf galaxies beyond the local Universe were nearly impossible to obtain. Samples of low-mass galaxies at higher redshift were often selected by their emission-line flux and therefore excluded galaxies without highly-elevated star formation. While such samples led to the detection of a significant number of low-mass galaxies with strong \ha\ emission at cosmic noon and thereby provided evidence of burstiness \citep{Maseda2013, Maseda2014, Atek2022}, these samples did not allow for a thorough study of the population characteristics. Made possible by JWST's unprecedented depth and the unique capabilities of the low resolution PRISM spectroscopy, our broadband-selected sample of low-mass galaxies at $z\approx 2$ represents a notable advance, enabling us to detect continuum emission in these systems and study the properties of the whole population, not only the most strongly star-forming galaxies. The sources in our sample span a broad range of SFRs, with many galaxies in the low-SFR tail as seen in \autoref{fig:sfms}.

Even at first glance, our unique sample provides insight into the behavior of this unexplored population, revealing a significant number of strong Balmer breaks. As we have shown above, this distribution constrains the nature of bursty star formation in these galaxies as it can only be reproduced if the galaxies are experiencing bursts which rise and fall on timescales of over $\sim$100 Myr and which fall to at least $\sim$0.8 dex below the SFMS. Though extreme-emission-line systems contain important information about the upper envelope of rising bursts in the low-mass population, our work demonstrates that observations of galaxies at the opposite extreme, those with the lowest star formation rates, are necessary for developing a complete picture of the population. 

Samples of low-mass broadband-selected galaxies such as ours present an exciting opportunity to empirically characterize the star-formation behavior of the cosmic noon population, providing a crucial bridge between well-characterized, but size-limited local samples \citep{Lee2009, Weisz2012, Emami2019} and highly incomplete but scientifically intriguing samples at very high redshift. Evidence has been presented that high-$z$ ($z\gtrsim 6$) galaxies do experience extended periods of dormancy \citep{Endsley2024, Dome2024, Looser2024, Looser2025, Trussler2025}, but the UV or emission-line-flux selections of high-$z$ samples make it nearly impossible to probe the length of these low-activity phases beyond $\sim50$ Myr. Considering that current observational capabilities do not allow for the construction of large, continuum-selected samples at such early epochs, additional work with larger, deeper, lower-mass samples at cosmic noon – the most distant epoch where this is currently possible – will provide useful benchmarks for studies of burstiness at high z.

\subsection{Long-timescale and High-amplitude Bursts in Dwarfs}

The distribution of Balmer break strengths observed in our broadband-selected low-mass sample of galaxies at cosmic noon can only be reproduced if the ensemble of galaxies experiences bursts of star formation that occur on timescales $\gtrsim100$ Myr and drop $\gtrsim 0.8$ dex below the main sequence. Are these preliminary conclusions consistent with previous results and what do they reveal about the drivers of star-formation bursts in this population? To investigate these questions, we first discuss existing frameworks connecting burst cycles to the physics of star formation.

In their analytical model explaining the origin of bursty star formation, \citet{Faucher-Giguere2018} conclude that bursty star formation should dominate in low-mass galaxies ($\lesssim 10^9 M_\odot$) at all redshifts and high-redshift galaxies ($z\gtrsim1$) at all masses. They argue that at low masses, star formation proceeds in a small number of gravitationally-bound clouds and so the globally-averaged star formation rate is governed by the stochasticity of those discrete regions. At high redshift, the free-fall time declines below the timescale for supernova explosions and so feedback cannot regulate gravitational collapse. In low-mass galaxies at high redshift, as we consider here, both of these factors likely contribute to the observed burstiness of the SFH. 

Building on this picture, \citet{Tacchella2020} present an extended regulator model, which describes how star formation is influenced by inflows, outflows, star-formation efficiency, and giant molecular cloud properties. Using a power spectrum density (PSD) formalism to quantify the degree of variability on different timescales, they demonstrate how each of the physical processes contributing to the global regulation of star formation influence the derived PSD. In brief, short time-scale variations are attributable to GMCs, with lifetimes on the order of $10$ Myr, while longer time-scale variations are regulated by gas cycling – i.e. the outflow and inflow of star-formation fuel. 

In the context of these physical models, our results indicate that burst cycles in low-mass galaxies at cosmic noon are likely driven primarily by global phenomena. Stochasticity of GMC formation and destruction cannot explain variations in SF with timescales of 100s of Myr; strong outflows driven by SN feedback and subsequent cooling and accretion of this material appears to be the dominant burst-regulating process. While our toy model cannot cannot capture variability on multiple timescales by design, future analysis will employ rigorous PSD-based models, allowing us to probe the relative contribution of various burst processes \citep[][E. Burnham et al, in prep]{Iyer2024}. Regardless, our analysis strongly suggests that the considered population is experiencing burst cycles with long timescales. Whether weaker, shorter-timescale variability also contributes is a question we must defer to later work. 

Due to the lack of comparable preexisting samples, there are few constraints on the burst parameters for galaxies with $\log M_\star/M_\odot<9$ at $z=1-3$, but evidence does exist for burstiness in this population. Recently, \citet{Clarke2024} compiled a sample of public JWST/NIRSpec observations spanning a range of $1<z<7$ down to masses of $M_\star\approx 10^8\,M_\odot$ and found the scatter in the SFMS with SFRs measured from \ha\ emission to be higher than the scatter in the SFMS with SFRs measured from UV emission at all masses for $z>1.4$, implying variability on timescales $\lesssim 200$ Myr. While this does not exclude the presence of long-timescale bursts, it does hint at an upper limit in SFR-variation timescales, or at least indicate that shorter-timescale variations may also contribute.

Locally, the strongest timescale constraints also come from \ha-UV comparisons. \citet{Emami2019} used data from the 11 Mpc \ha\ and UV Galaxy Survey \citep[11HUGS,][]{Kennicutt2008, Lee2009} to constrain burst timescales and amplitudes using a model similar to ours: a periodic exponential increase and decay. They separate their sample into mass bins and find burst duration generally increasing with mass from $\lesssim 100$ Myr for galaxies with $\log M_\star/M_\odot <7$ to $\sim700$ Myr for galaxies with $ 9<\log M_\star/M_\odot <10$. In their central mass bin of  $8<\log M_\star/M_\odot <8.5$, they find a duration of $\sim 200$ Myr, consistent with our findings. While our sample is too small to subdivide into bins of mass or redshift, the \citet{Emami2019} analysis highlights that such specification could lead to additional insights. This approach will be made possible by a larger sample. Regardless, the general agreement between our conclusions and these local dwarf galaxy results indicates that the dominant burst drivers may not have changed significantly since cosmic noon, at least not for galaxies in the bright-dwarf mass range. It appears that the stochasticity of GMC formation and destruction becomes more significant at lower masses. A larger dynamic range is required to probe this transition at higher $z$.

Notably, the constraints discussed above were derived from comparisons of \ha\ and UV emission in dwarf galaxies. Because UV emission traces star formation averaged over $\sim100$ Myr, the SFRs derived from \ha\ and UV emission will equilibrate after a few 100 Myr of constant star formation and so cannot be reliably used to probe variability on much longer timescales. Because the Balmer break tracks stellar age on longer timescales than the UV emission (on the order of main-sequence lifetimes of low-mass A stars -- i.e. up to $\sim1$ Gyr), it is a useful diagnostic for identifying long term variations in star formation, beyond what is possible with UV and \ha\ alone. Planned observations (P.I. Whitaker \#17730) will obtain HST imaging of the UNCOVER field in the rest-frame FUV, allowing for analysis based on all three observables, which will provide powerful constraints over a wide range of burst timescales.

\subsection{Towards a Unified Perspective on Low-mass Galaxies}\label{subsec:sfms}

Analysis of star formation behavior in populations of galaxies typically begins with a separation of quiescent and star-forming sources, motivated largely by observed bimodality in large photometric samples of galaxies \citep{Strateva2001}. The SFMS and its scatter are typically fit only to the ``star-forming" sources in a sample, often identified by some \haew, \ha\ flux, or sSFR threshold. In recent years, many have advocated for an updated perspective on the SFMS, one that models the density of the entire unified population without imposing a star-formation selection \citep{Feldmann2017a, Leja2022}. These results indicate that galaxies in the SFR-$M_\star$ plane (at least at $\log M_\star>10^9\,M_\odot$ at $z\lesssim1$) are better described by their density distribution than by a linear fit and intrinsic scatter, the latter being especially sensitive to the sample selection. Our results provide additional support for this updated perspective and demonstrate the necessity of viewing galaxy populations as unified rather than bifurcated, especially at low mass and high $z$.

If, as our results indicate, low-mass galaxies at cosmic noon spend reasonably long periods below the main sequence, subdividing similar samples based on current rate of star formation could obfuscate this bursting behavior. \textcolor{black}{While some number of galaxies below the SFMS are likely undergoing permanent quenching, recent results suggest that this does not begin in earnest until $z\sim1.5$ \citep{Cutler2025}. If they were present in our sample, we would expect old, quiescent galaxies to exhibit 4000\,\AA\ breaks, which we do not observe (see Figures~\ref{fig:datasmall}, \ref{fig:databig1} \ref{fig:databig2})}. Only a single galaxy in our sample would be classified as quiescent by classical UVJ \citep{Whitaker2011} or NUVJr \citep{Ilbert2013} metrics (ID \#21055), but as shown in \autoref{fig:sfms} our sources span a broad sSFR range -- nearly 2 dex -- though we note that we are incomplete to sources with dimmer continua, weaker \ha, and stronger breaks. 

The absence of these galaxies does not impact our conclusion that star formation burstiness at cosmic noon must have \db\ $\gtrsim 100$ Myr; in fact, the finding that a significant portion of the galaxy population is found to be in a lull of star formation is made even stronger by the knowledge that we are incomplete only to the galaxies that are in that lull. 
Deeper photometric data would reveal even less-active galaxies and further populate the low-mass, low-SFR region of the SFR-$M_\star$ plane. We caution others using similarly deep samples to avoid imposing SFR cuts and to consider the framework of bursty star formation when analyzing low-mass samples.

Despite the simplicity of our model SFHs and initial statistical analysis, our simulated model SFHs do appear to accurately replicate the low-SFR tail of the observed sample, as shown in \autoref{fig:sfms}. While the shape and uniformity of the burst phase is poorly matched to the data (likely the bursts should be less peaked and more extended), the depth and duration of the dormant phase is representative of the spread in the PRISM and parent samples. What we have shown through our analysis of the Balmer break distribution is also evident in the sSFR-$M_\star$ plane: a significant number of low-mass galaxies at cosmic-noon dwell below the main sequence. The Balmer break analysis further demonstrates that this SFR distribution is driven by dynamic movement around the MS, not by a time-constant spread in SFR among galaxies of a given stellar mass.

Future work will extend these results further, with photometric measurements allowing us to probe even less-massive galaxies with even lower rates of star formation. Though the precision of the presented constraints are limited by the small sample size and the specific parameterization of the bursty SFHs, the large photometric sample will enable us to repeat this analysis more generically and on a much larger sample. This analysis will require fewer simplifying assumptions about the uniformity of bursts within the ensemble and allow us to explore constraints on predicted main-sequence scatter that was not possible within our current framework. The UNCOVER and MegaScience programs provide deep 20-band photometry (5$\sigma$ depth of 29 mag in F210M) of a 30 arcmin$^2$ region centered on A2744, including imaging in 12 medium bands. The broad wavelength coverage and inclusion of the medium bands allows us to measure reliable photometric redshifts. Spectral features such as \ha\ emission and Balmer break strength should be accurately recoverable, albeit with lower precision. Having established the power of these metrics to probe bursts, we will leverage the deep and well-sampled photometry to better constrain the burst parameters in smaller bins of mass and redshift.

\section{Summary}

We use a novel broadband-selected sample of JWST PRISM spectroscopy of low-mass galaxies at $z=1-3$ to study the burstiness of star formation at cosmic noon. The low-resolution spectra, which enable continuum detection in these dim sources, reveal numerous strong Balmer breaks, indicating extended periods of suppressed star formation in a number of the sources. Under the simplifying assumption that the galaxies in this mass and redshift range share a common burst timescale and amplitude, we generate synthetic observations using a toy parameterization of bursty star formation histories. The simulated samples exclude regions of the timescale-amplitude parameter space, demonstrating that the SF-bursts must occur on long timescales ($\gtrsim 100$ Myr) and drop well below ($\gtrsim 0.8$ dex) the SFMS. 

Our conclusions are consistent with previous constraints on burst timescales ($\sim200$ Myr), but support an updated perspective on low-mass galaxies that considers star-forming and dormant sources as a unified population. The demonstrated prevalence of long timescale bursts indicates that galaxy-scale processes (e.g. Baryon cycling via feedback-driven outflows and inflows) are likely the dominant regulator of star formation in these systems. Future work will extend this analysis to the full MegaScience 20-band photometric sample, applying a more sophisticated analytical approach to remove the dependence on the SFH parameterization and marginalize over sources of uncertainty such as dust, metallicity, and SPS model choices. 
\vspace{1in}
\section*{Acknowledgments}
A.M. acknowledges support from the National Science Foundation Graduate Research Fellowship under Grant No. 2039656. A.M. thanks Helena Treiber for helpful discussions. Support for this work was provided by The Brinson Foundation through a Brinson Prize Fellowship grant. 

This work is based in part on observations made with the NASA/ESA/CSA James Webb Space Telescope obtained from the Space Telescope Science Institute, which is operated by the Association of Universities for Research in Astronomy, Inc., under NASA contract NAS 5–26555. Financial support for programs JWST-GO-1837, JWST-GO-2561, JWST-GO-4111, and JWST-GO-6405 is gratefully acknowledged and is provided by NASA through grants from the Space Telescope Science Institute, which is operated by the Associations of Universities for Research in Astronomy, Incorporated, under NASA contract NAS 5-03127. These observations are associated with programs JWST-GO-2561, JWST-ERS-1324, JWST-DD-2756, JWST-GO-4111, HST-GO-11689, HST-GO-13386, HST-GO/DD-13495, HST-GO-13389, HST-GO-15117, and HST-GO/DD-17231. Some of the data presented in this article were obtained from the Mikulski Archive for Space Telescopes (MAST) at the Space Telescope Science Institute. The specific observations analyzed can be accessed via \dataset[10.17909/grtw-hr67]{https:///doi.org/10.17909/grtw-hr67}. Some of the data products presented herein were retrieved from the Dawn JWST Archive (DJA). DJA is an initiative of the Cosmic Dawn Center, which is funded by the Danish National Research Foundation under grant No. 140.

\textit{Facilities}: JWST (NIRCam and NIRSpec), HST (WFC3 and ACS)

\textit{Software}: \texttt{NumPy} \citep[\url{https://numpy.org},][]{Harris2020}, \texttt{Astropy}  \citep[\url{https://astropy.org},][]{Astropy2013, Astropy2018, Astropy2022}, \texttt{Matplotlib} \citep[\url{https://matplotlib.org},][]{Hunter2007}, \texttt{SciPy} \citep[\url{https://scipy.org},][]{Virtanen2020}.

\bibliography{bursty}

\appendix
\begin{figure*}
    \centering
\includegraphics[width=1\linewidth]{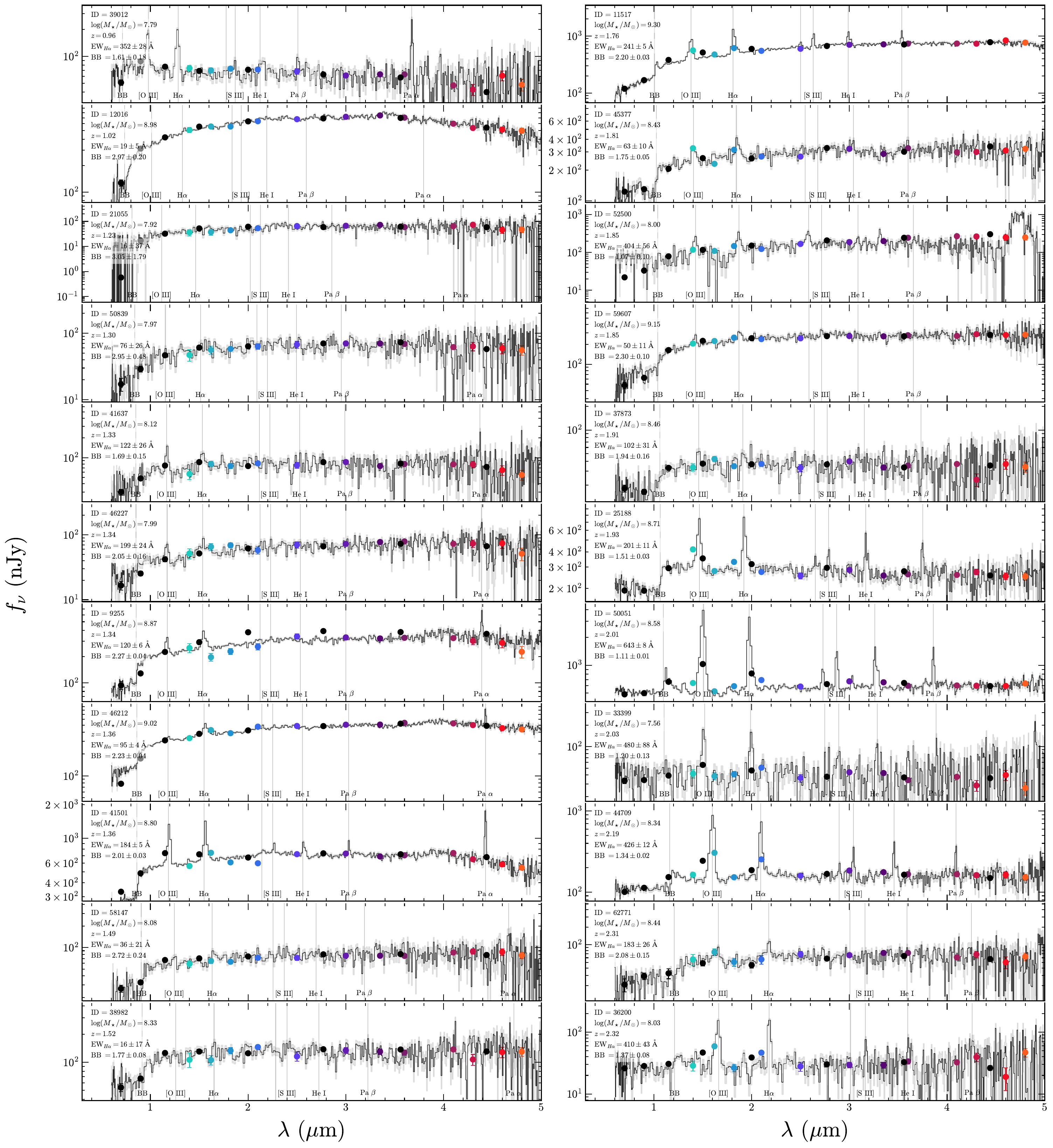}
    \caption{NIRSpec/PRISM spectra for half of the 43 galaxies in our sample sorted by increasing redshift. The black line shows the spectral flux density $f_\nu$ and the shaded region shows the $1\sigma$ uncertainty. Photometry from UNCOVER and MegaScience are plotted over the spectra with broadbands in black and medium bands colored according to the filters shown in \autoref{fig:datasmall}. Vertical lines highlight selected spectral features including the Balmer break (BB) and nebular emission lines. The source ID, stellar mass, spectroscopic redshift, \haew\, and Balmer break strength are shown for each object. The galaxies were selected for observation based on F200W magnitude and photometric redshift. Only galaxies with SED-fitted stellar mass below $10^{9.5}\,M_\odot$ were included in the analyzed sample.  }
    \label{fig:databig1}
\end{figure*}

\begin{figure*}
    \centering
\includegraphics[width=1\linewidth]{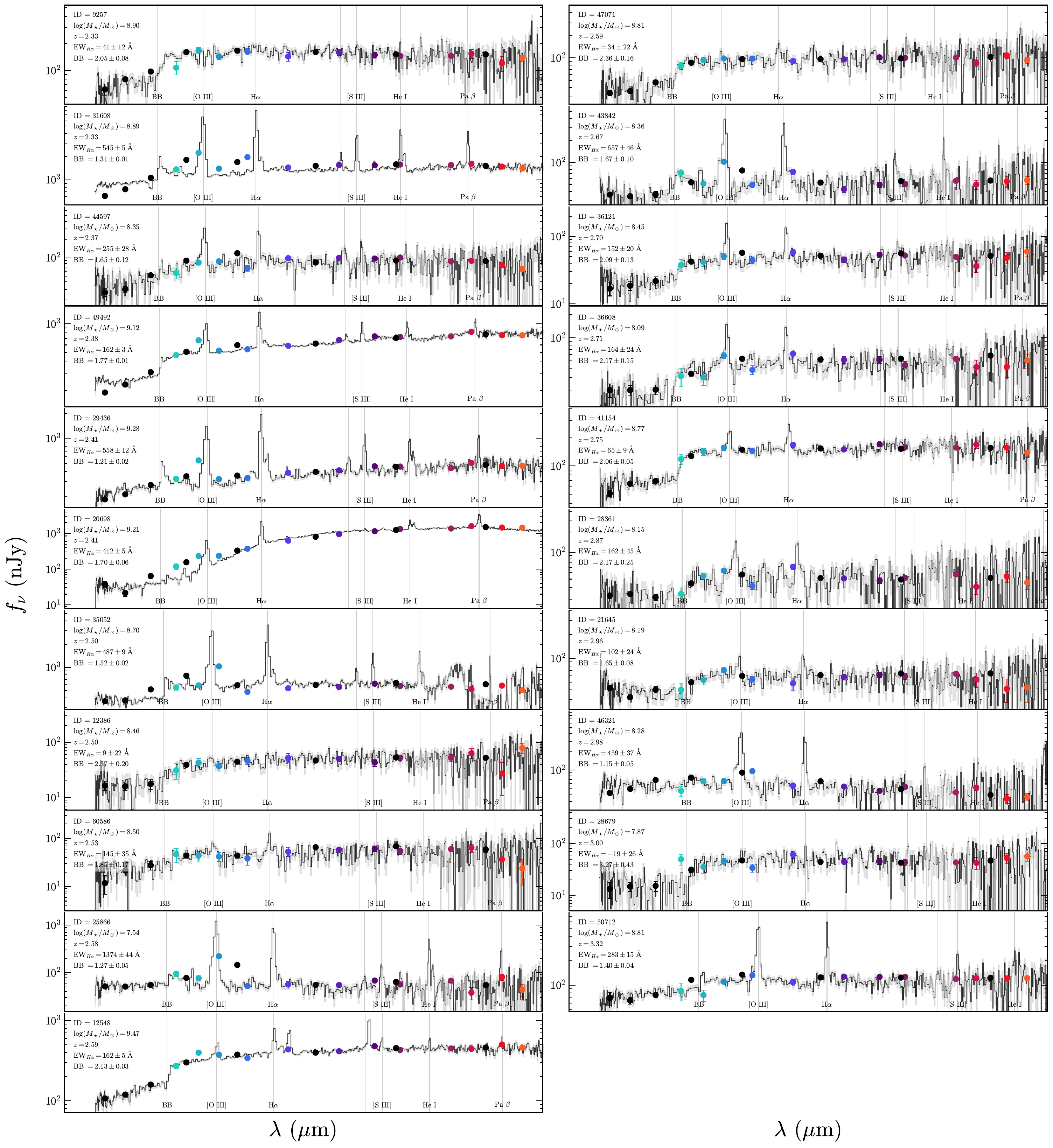}
    \caption{The same as \autoref{fig:databig1}, but for the remaining objects. }
    \label{fig:databig2}
\end{figure*}

\section{Full sample of PRISM spectra}\label{app:data}
We present the full sample of 43 PRISM spectra in \autoref{fig:databig1} and \autoref{fig:databig2}.

\section{Effects of metallicity and  dust}\label{app:dust}

\begin{figure}
    \centering
\includegraphics[width=1\linewidth]{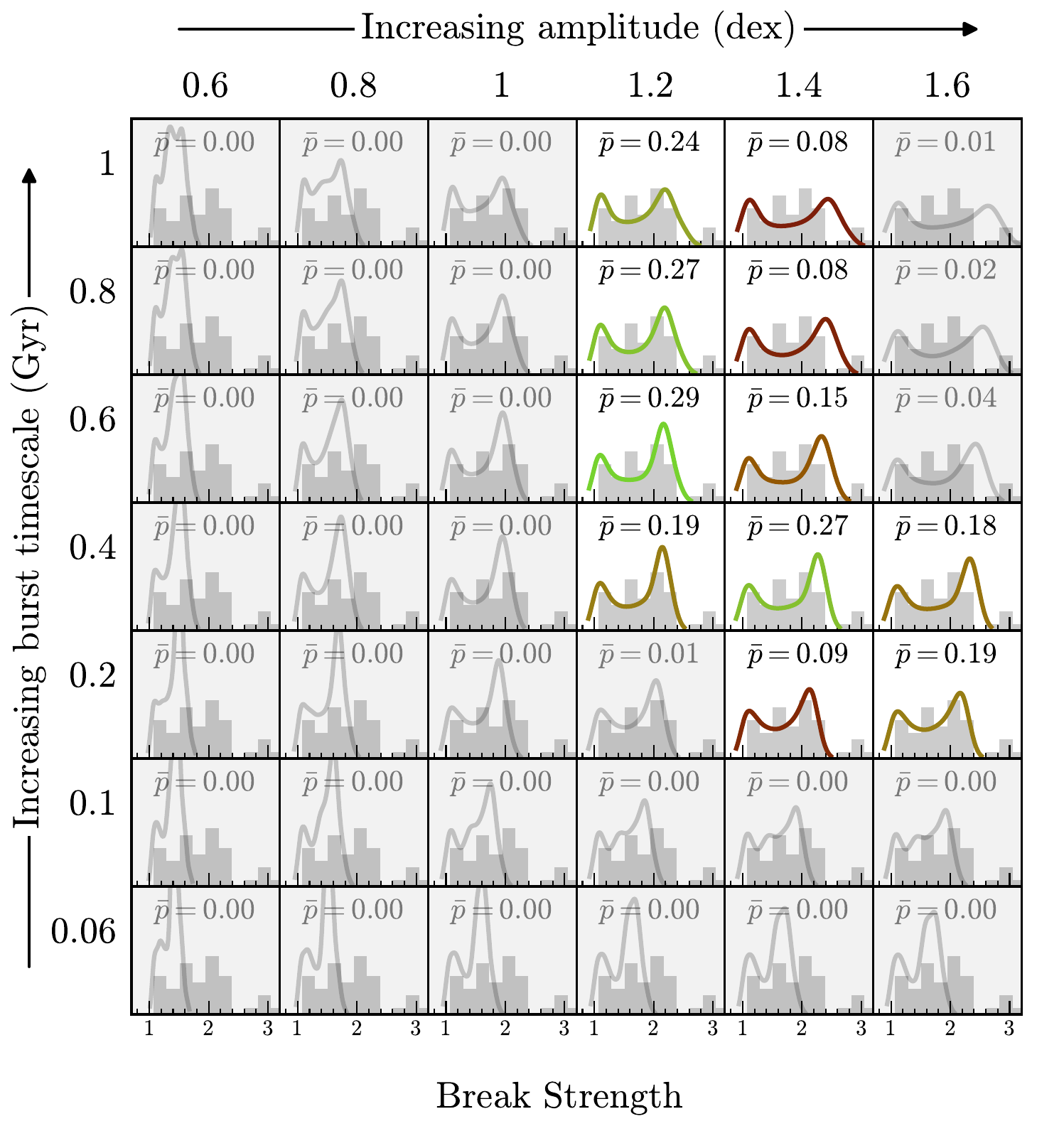}
    \caption{Same as \autoref{fig:bb_hist}, but showing Balmer break distributions from models generated with no dust and metallicity fixed to $Z=0.1\,Z_\odot$.}
    \label{fig:nodust}
\end{figure}

As described in Section~\ref{subsec:modelgen}, when generating the simulated galaxy populations and spectroscopy, we draw the gas-phase and stellar metallicity as well as the optical dust of diffuse gas ($\tau$) from the distribution of SED-fitted parameters as measured in \citet{Wang2024}. While this approach likely represents a realistic spread in metallicity and attenuation, the metallicities and $\tau$ values measured from the SED-fits are not very well constrained. To further investigate the dependence of our results on these choices, we repeat our analysis with different assignments of metallicity and dust attenuation. 

\subsection{Dust}
We generate three sets of models, each with a single value of $A_V$ for all simulated galaxies in the model population: one with $A_V=0$ mag (dust-free), one with $A_V=0.2$ mag (close to the median $A_V=$ 0.21 mag from the SED fits to the photometry), and one with $A_V=0.4$ mag as an extreme case. For these models, we fix $Z=0.1\,Z_\odot$ to better isolate the impact of dust. We note that the difference in the preferred burst parameter space is minimal when comparing models with SED-sampled $\tau$ and SED-sampled metallicity to models with SED-sampled $\tau$ and fixed $Z=0.1\,Z_\odot$. 

With $A_V=0$ mag, the preferred region of burst parameter space shifts to higher amplitude (\sigsf\ $\gtrsim1.2$ dex) and longer timescale (\db\ $\gtrsim$ 200 Myr) as shown in \autoref{fig:nodust}. Without attenuation increasing the measured break strength, longer dormant periods and more extreme departures below the SFMS are required to match the observed distribution. With $A_V=0.2$, the allowed regions of parameter space are nearly identical to our default model with sampled $\tau$, but with slightly higher lower-amplitude limits (\sigsf\ $\gtrsim1$ dex). With $A_V=0.4$ mag, the simulated Balmer break distributions are always inconsistent with the observed distribution; due to the reddened continuum slope, all of the Balmer breaks are made to appear stronger and there are no weak breaks measured from the generated models. 

It is well established that low-mass galaxies are generally less dusty than their massive counterparts, with lower dust-to-gas ratios and lower attenuation. This lack of significant attenuation is also obvious from the PRISM spectra themselves, the majority of which exhibit no signs of significant reddening (see Figures~\ref{fig:datasmall}, \ref{fig:databig1}, and \ref{fig:databig2}). The $A_V=0.4$ case is therefore unrealistic and the experiment provides further evidence of the low-dust content of the population.

The dust-free models disallow all values of \db\ $\lesssim200$ Myr and \sigsf\ $\lesssim1.2$ dex while the lightly-attenuated models disallow all values \db\ $\lesssim100$ Myr and \sigsf\ $\lesssim1$ dex. We conclude that our results are robust to realistic changes in attenuation; the strong Balmer breaks in the galaxies cannot be explained by dust alone. Extended periods of dormancy are still required to match the observed distribution. Exactly how long ($\sim100$ or $\sim200$ Myr) and how dormant ($0.8$ or 1 dex below the SFMS) these periods are will be better constrained by future analysis of larger samples.

\subsection{Metallicity}
Next, we explore the effect of changing metallicity. Increasing metallicity generally increases the apparent strength of the Balmer break, due to a combination of cooler stellar atmospheres, accelerated stellar evolution, and metal-line blanketing, which impact the continuum shape and strength of the Balmer break itself. The effect of changing metallicity is more complicated, but overall less significant than the impact of changing dust attenuation; in addition to shifting to stronger breaks, the width of the generated distribution of Balmer breaks also increases with metallicity. We conduct another brief experiment as above, repeating the model generation (without dust) with the metallicity fixed at $Z=0.1\,Z_\odot$, $Z=0.3\,Z_\odot$ ($\log Z/Z_\odot = -0.5$) and $Z=Z_\odot$. The models with $Z=0.1\,Z_\odot$ yield distributions that are nearly identical to those with metallicities sampled from the SED-fitted distribution (with no attenuation). The fixed 10\%\,$Z_\odot$ case prefers models with \db\ $\gtrsim200$ Myr and \sigsf\ $\gtrsim$ 1.2 dex while the sampled-metallicity case yields $\gtrsim$ 1 dex and identical limits for \db. At $Z=0.3\,Z_\odot$, there is also little difference in the distributions, with the preferred and disallowed regions of parameter space largely unchanged. Only at  $Z=Z_\odot$ do we find the break strengths to be significantly affected and populations generated with shorter timescales (\db\ $\gtrsim 60$ Myr) and smaller amplitude (\sigsf\ $\gtrsim 0.8$) become consistent with the data. While it is possible that some small number of the galaxies in our sample have metallicities as high as $Z_\odot$, it is highly unlikely that the entire set of low-mass galaxies at cosmic noon would be so metal rich. Considering this context, and the fact that the maximum metallicity from the SED-fitted values is $Z\approx0.6\,Z_\odot$, we conclude that our results are robust to reasonable increases in gas-phase and stellar metallicity.

Bursty star formation is still needed to explain the observed sample properties. Planned future extensions of this work to a much larger, photometric sample will incorporate more sophisticated analyses and include marginalization over complicating factors such as metallicity and attenuation.

\section{Bootstrap resampling}\label{app:bootstrap}

\begin{figure}
    \centering
\includegraphics[width=1\linewidth]{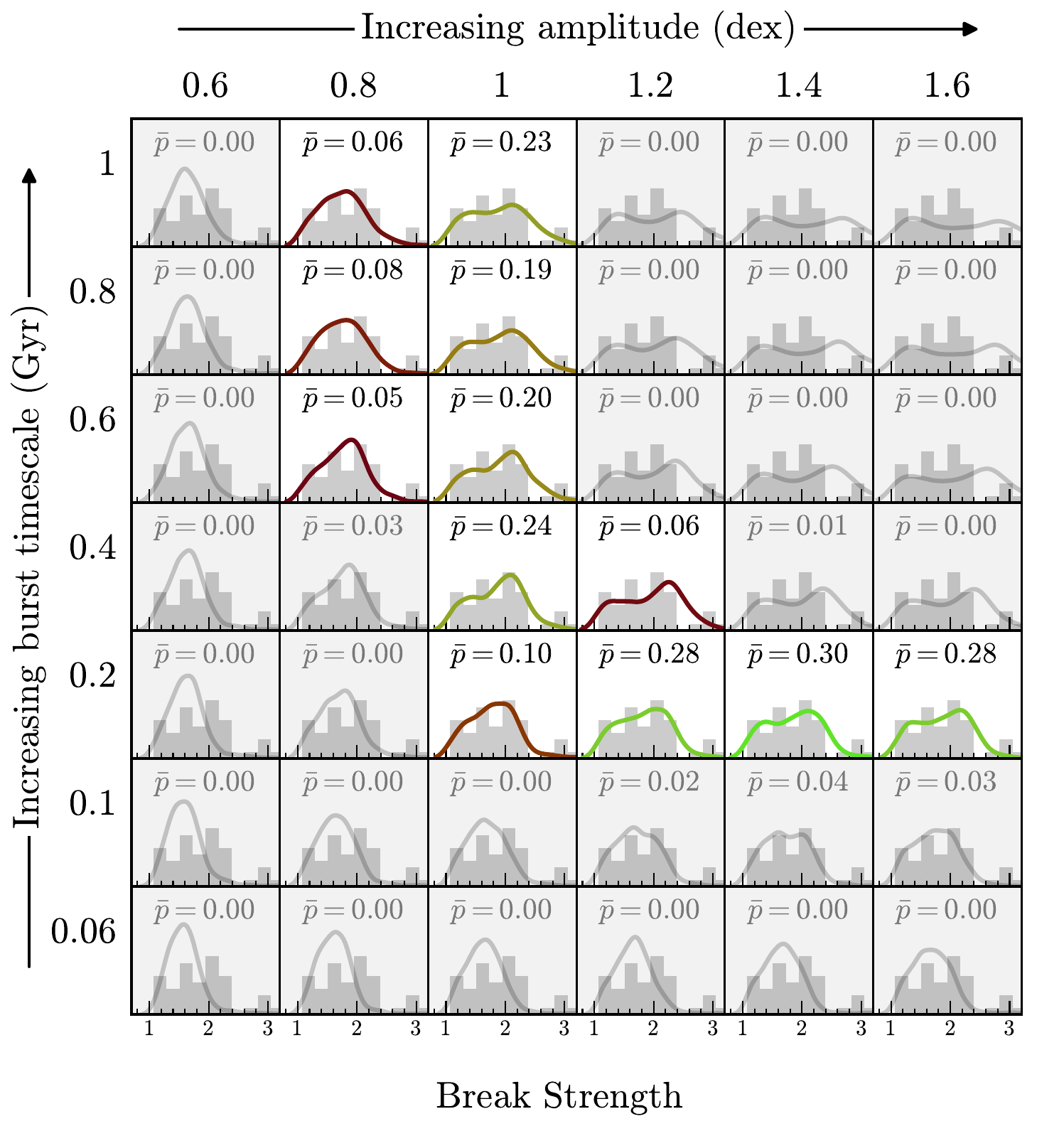}
    \caption{Same as \autoref{fig:bb_hist}, but showing median p-values from bootstrap resampling. The results are consistent with what is reported in the main text in \autoref{fig:bb_hist}.}
    \label{fig:histgrid_bootstrap}
\end{figure}

Although the size of our spectroscopic sample is small, the presence of strong Balmer breaks is striking. In particular, we see Balmer breaks with strengths greater than 2 in nearly half of the galaxies. We are therefore confident that our conclusions are driven by the overall demographics of the sample and not by a small number of anomalous sources. To demonstrate this, we repeat the analysis performed in Section~\ref{subsec:hists} with bootstrap resampling. For each realization, we sample with replacement from the galaxies in the sample (also perturbing their best-fit values to account for measurement uncertainty as described in Section~\ref{subsec:hists}) and measure the KS-test p-value comparing the bootstrapped sample to the model-distribution for each \db--\sigsf\ pair. As before, we take the median p-value from the 1000 iterations and report the values in \autoref{fig:histgrid_bootstrap}. The preferred and excluded burst parameters are nearly identical to those we found in Section~\ref{subsec:hists} and if anything, indicate a longer lower limit (closer to 200 Myr) for the burst timescale \db.

Despite the small size of our sample, our results robustly indicate that the number of strong Balmer breaks can constrain the burst behavior of the galaxies in this population. Larger samples and more sophisticated modeling will be necessary for specifying the timescale constraints in the context of realistic star formation histories.

\end{document}